\PassOptionsToPackage{table}{xcolor}
\documentclass[10pt,twoside]{article}
\usepackage{nca-sample}
\usepackage{multirow,colortbl,textcomp,bbm,enumitem,wrapfig,etoolbox,placeins,amsmath}
\usepackage[english]{babel}
\usepackage[autostyle,english=american]{csquotes}
\usepackage[algo2e,ruled]{algorithm2e}
\usepackage{adjustbox}
\usepackage{subcaption}
\usepackage{tikz}
\usetikzlibrary{positioning,arrows.meta}
\BeforeBeginEnvironment{tabular}{\begin{adjustbox}{max width=\linewidth}}
\AfterEndEnvironment{tabular}{\end{adjustbox}}
\definecolor{lightgray}{gray}{0.85}
\definecolor{lightgreen}{RGB}{220,255,220}
\definecolor{lightorange}{RGB}{255,245,220}
\definecolor{lightblue}{RGB}{220,240,255}
\MakeOuterQuote{"}
\AtBeginEnvironment{table}{\footnotesize}

\RestyleAlgo{ruled}
\DontPrintSemicolon
\SetAlFnt{\footnotesize}
\SetAlCapFnt{\normalsize}
\SetAlCapNameFnt{\normalsize}
\newcommand{\NumSegOverRedLo}{1.2}
\newcommand{\NumSegOverRedHi}{2.7}
\newcommand{\NumSegRecCostLo}{1.3}
\newcommand{\NumSegRecCostHi}{5.7}

\newcommand{\NumFullPadLo}{25\%}
\newcommand{\NumFullPadHi}{85\%}
\newcommand{\NumSegFPRSAFrom}{5.4\%}
\newcommand{\NumSegFPRSATo}{2.7\%}
\newcommand{\NumSegFPRLingFrom}{26.9\%}
\newcommand{\NumSegFPRLingTo}{9.9\%}
\newcommand{\NumSegFPRCEASFrom}{1.4\%}
\newcommand{\NumSegFPRCEASTo}{1.2\%}
\newcommand{\NumSegPAUCDiff}{0.002}

\newcommand{\NumSegRetLo}{83\%}
\newcommand{\NumSegRetHi}{94\%}
\newcommand{\NumFullRetLo}{31\%}
\newcommand{\NumFullRetHi}{90\%}
\newcommand{\NumLlamaOneLo}{83.7\%}

\newcommand{\NumCCOneLo}{71.4\%}
\newcommand{\NumCCOneHi}{73.2\%}
\newcommand{\NumLlamaDefaultFPR}{0.2\%}
\newcommand{\NumLlamaDefaultRecall}{83.7\%}
\newcommand{\NumLatLocal}{0.20}
\newcommand{\NumLatCloud}{0.33}
\newcommand{\NumLatLLM}{4.6}
\newcommand{\NumSpeedLo}{14}
\newcommand{\NumSpeedHi}{24}
\newcommand{\NumCostCloud}{6.45}
\newcommand{\NumCostLLM}{64}

\newcommand{\NumAuthN}{180}
\newcommand{\NumAuthPhishN}{140}
\newcommand{\NumAuthStrict}{37}
\newcommand{\NumAuthRfc}{13}
\newcommand{\NumAuthRedPct}{65\%}
\newcommand{\NumAuthPhishKept}{140}
\newcommand{\NumAuthDead}{7}
\newcommand{\NumRevSeedMax}{2.1\%}
\newcommand{\NumPhaseOneOverride}{0.1\%}
\newcommand{\NumWoneConfigs}{20}
\newcommand{\NumWoneFacLo}{1.9}
\newcommand{\NumWoneFacHi}{3.8}
\newcommand{\NumWoneDefaultRank}{9}
\newcommand{\NumToneBminLo}{0.28}
\newcommand{\NumToneBminHi}{0.43\%}
\newcommand{\NumToneBaseLo}{15}
\newcommand{\NumToneBaseHi}{151}
\newcommand{\NumToneCCLo}{3.2}
\newcommand{\NumToneCCHi}{57}
\newcommand{\NumToneLenLo}{1.2}
\newcommand{\NumToneLenHi}{1.8}
\newcommand{\NumToneNvalLo}{240}
\newcommand{\NumToneNvalHi}{409}
\newcommand{\NumCsixFacLo}{5.5}
\newcommand{\NumCsixFacHi}{21}
\newcommand{\NumCsixRecLossLo}{11}
\newcommand{\NumCsixRecLossHi}{26}
\newcommand{\NumCsixBTenLo}{8.1}
\newcommand{\NumCsixBTenHi}{92.3\%}
\newcommand{\NumVoneFPRLo}{2.7}
\newcommand{\NumVoneFPRHi}{4.7\%}
\newcommand{\NumVoneLingFrom}{26.9\%}
\newcommand{\NumVoneLingTo}{4.7\%}
\newcommand{\NumVoneEscLo}{14}
\newcommand{\NumVoneEscHi}{54\%}
\newcommand{\NumVoneLLMLo}{6.0}
\newcommand{\NumVoneLLMHi}{8.4\%}
\newcommand{\NumBtwoOverPairs}{4}
\newcommand{\NumBtwoBaseLo}{7.6}
\newcommand{\NumBtwoBaseHi}{58.4\%}
\newcommand{\NumBtwoCCLo}{0.0}
\newcommand{\NumBtwoCCHi}{16.0\%}
\newcommand{\NumBtwoLtwoBaseLo}{0.2}
\newcommand{\NumBtwoLtwoBaseHi}{3.4\%}
\newcommand{\NumBtwoLtwoCCLo}{1.7}
\newcommand{\NumBtwoLtwoCCHi}{3.9\%}
\newcommand{\NumRsixCCFPR}{0.3\%}
\newcommand{\NumRsixCCRecall}{62.0\%}
\newcommand{\NumRsixLoneLo}{5.5}
\newcommand{\NumRsixLoneHi}{8.5\%}
\newcommand{\NumRsixBaseLo}{5.0}
\newcommand{\NumRsixBaseHi}{10.0\%}
\newcommand{\NumFrCCOne}{73.2\%}
\newcommand{\NumFrLLMOne}{56.9\%}
\newcommand{\NumFrCCTen}{90.4\%}
\newcommand{\NumFrLLMTen}{96.6\%}
\newcommand{\NumFrGapOne}{16}
\newcommand{\NumFrBTenAUROC}{0.989}
\newcommand{\NumFrCCpAUCLo}{0.850}
\newcommand{\NumFrCCpAUCHi}{0.864}
\newcommand{\NumLeakRows}{2,211}
\newcommand{\NumLeakTotal}{2,300}
\newcommand{\NumLeakPct}{96\%}
\newcommand{\NumOldRecall}{99.1\%}
\newcommand{\NumOldFPR}{4.1\%}
\newcommand{\NumNazOverlapPos}{602}
\newcommand{\NumDPFamilies}{3{,}075}
\newcommand{\NumDPGroups}{2,699}
\newcommand{\NumShiftCCFPR}{5.4\%}
\newcommand{\NumShiftCCRecall}{86.4\%}
\newcommand{\NumShiftBaseLo}{20.2}
\newcommand{\NumShiftBaseHi}{22.9\%}
\newcommand{\NumShiftZeroFPBest}{8.9\%}
\newcommand{\NumShiftCCZeroFP}{1.2\%}
\newcommand{\NumShiftCCZeroFPRecall}{77.3\%}

\newcommand{\NumBElevenRecall}{79.8\%}
\newcommand{\NumBElevenZeroFPRecall}{25.8\%}
\newcommand{\NumUnwantedCCFPR}{0.1\%}
\newcommand{\NumUnwantedCCRecall}{14.6\%}
\newcommand{\NumLLMFPR}{0.7\%}

\newcommand{\NumLLMCalls}{10,527}
\newcommand{\NumLLMFailures}{0}
\newcommand{\NumLLMCostPerK}{\$0.033}
\newcommand{\NumLLMCostTotal}{\$0.35}

\newcommand{\NumRtwoCeasCC}{1.4\%}

\newcommand{\NumRtwoLingCC}{26.9\%}

\newcommand{\NumRtwoFacLo}{2.7}
\newcommand{\NumRtwoFacHi}{6.1}
\newcommand{\NumRtwoBaseMinOver}{9}
\newcommand{\NumRtwoBaseMaxOver}{94}
\newcommand{\NumRfiveCCRecall}{87.7\%}
\newcommand{\NumRfiveBElevenRecall}{27.2\%}
\newcommand{\NumRfiveMaxFPR}{1.2\%}
\newcommand{\NumLLMLingFPR}{1.3\%}
\newcommand{\NumRfourMaxGain}{0.6}
\newcommand{\NumOntRules}{8}
\newcommand{\NumChallengeWrong}{30}
\newcommand{\NumChallengeVoneSim}{5}
\newcommand{\NumChallengeVtwo}{0}
\newcommand{\NumChallengeRight}{9 / 9}
\newcommand{\NumPrivRows}{26,075}

\newcommand{\NumPrivResidualIP}{690}
\newcommand{\NumPrivResidualIPPct}{2.6\%}
\newcommand{\NumPrivResidualDates}{22}
\newcommand{\NumPlantN}{4,000}
\newcommand{\NumPlantLost}{63}
\newcommand{\NumPlantNameCued}{94.4\%}
\newcommand{\NumPlantNameUncued}{79.8\%}

\newcommand{\NumPlantNameLo}{57\%}

\newcommand{\NumPlantNameHi}{96\%}
\newcommand{\NumPlantDobDMY}{0\%}
\newcommand{\NumPlantDobOther}{100\%}

\newcommand{\CCTitle}{CyberCane: a privacy-bounded neuro-symbolic retrieval framework for phishing detection under distribution shift}

\newcommand{\CCFunding}{The authors did not receive support from any organization for the submitted work.}

\renewcommand{\ArticleTitle}{\CCTitle}
\renewcommand{\ArticleAuthors}{Safayat Bin Hakim \textperiodcentered\ Aniqa Afzal \textperiodcentered\ Qi Zhao \textperiodcentered\\Vigna Majmundar \textperiodcentered\ Pawel Sloboda \textperiodcentered\ Houbing Herbert Song}
\renewcommand{\ArticleDates}{University of Maryland, Baltimore County; Bowie State University}
\renewcommand{\ArticleNotice}{Author contact information appears before the references.}
\renewcommand{\ArticleIdentifier}{CyberCane}
\renewcommand{\JournalName}{Preprint}
\renewcommand{\ArticleType}{}
\renewcommand{\ArticleNumber}{}
\renewcommand{\FooterMarker}{}
\renewcommand{\OpeningCitation}{}
\renewcommand{\OpeningStatus}{}
\hypersetup{pdftitle={CyberCane: a privacy-bounded neuro-symbolic retrieval framework for phishing detection under distribution shift},pdfauthor={Safayat Bin Hakim; Aniqa Afzal; Qi Zhao; Vigna Majmundar; Pawel Sloboda; Houbing Herbert Song},pdfsubject={Phishing detection under distribution shift}}
\begin{document}
\raggedbottom

\MakeArticleTitle

\begin{ArticleAbstract}
Organizations usually calibrate phishing detectors on public corpora, because privacy rules make their own labeled legitimate mail hard to collect. This practice breaks the false-positive budget of trained detectors: calibrated for a 1\% false-positive rate (FPR) on older public mail and deployed on DataPhish 2025, they realize \NumShiftBaseLo--\NumShiftBaseHi{} FPR, \NumRtwoBaseMinOver--\NumRtwoBaseMaxOver$\times$ the budget across three calibration sources. The cause is calibration transfer, not ranking: the classifiers separate the deployment mail well, but their thresholds do not transfer. We introduce CyberCane, a privacy-bounded neuro-symbolic detector that scores email only by similarity to known phishing in a redacted, training-only index, plus deterministic rules, and calibrates verdicts to an explicit FPR budget. Because legitimate mail never enters its score, only its threshold depends on the calibration source. Under the same shift CyberCane realizes \NumShiftCCFPR{} FPR at \NumShiftCCRecall{} recall, reducing the overrun \NumRtwoFacLo--\NumRtwoFacHi$\times$ across sources, seeds, a second shift family, and a local embedder. A segment-level variant, scoring an email's most phishing-like part, lowers the realized FPR a further \NumSegOverRedLo--\NumSegOverRedHi$\times$ at equal ranking quality and withstands benign padding. CyberCane screens an email locally in \NumLatLocal{}\,s, sending no text out; a 70B zero-shot LLM ranks better at strict FPR but is \NumSpeedLo--\NumSpeedHi$\times$ slower and must receive each email. We also show that LLM rewriting does not evade embedding-based detectors and that a random split of DataPhish 2025 leaks \NumLeakPct{} of its test set. All protocols were fixed in advance, and the released splits and code let a deployer measure realized FPR before trusting a budget.
\end{ArticleAbstract}

\ArticleKeywords{Phishing detection \textperiodcentered\ distribution shift \textperiodcentered\ false-positive control \textperiodcentered\ retrieval-augmented detection \textperiodcentered\ neuro-symbolic AI \textperiodcentered\ privacy}

\FloatBarrier
\section{Introduction}\label{sec:introduction}

Phishing remains a primary initial-access vector across industries, and the human element contributes to over 60\% of data breaches~\citep{verizon2025dbir}. Privacy-critical organizations face a particularly sharp version of the problem. In healthcare, a false alarm can delay an appointment reminder or a prescription notification, so the tolerable false-positive rate (FPR) is small and explicit; older adults are measurably worse at distinguishing phishing from legitimate mail~\citep{pehlivanoglu2024phishing,gallo2024human}; and regulations such as the US health-privacy law HIPAA constrain which email content may be sent to third-party services~\citep{shanmugarasa2025sok}. Generative AI adds pressure by making convincing, domain-specific phishing cheap to produce~\citep{dipalma2025airisks,heiding2024devising}.

These constraints meet in a step that detection papers rarely examine: \emph{calibration}. A deploying organization sets an operating threshold so that at most a fixed fraction of legitimate mail is flagged. Doing so requires labeled legitimate mail, which is exactly the data privacy rules make hard to collect and share. In practice, thresholds are therefore set on public corpora whose legitimate mail differs from the organization's own. Learned detectors reach near-ceiling accuracy on held-out splits of such corpora, and our leak-free experiments reproduce this on the classic Nazario and SpamAssassin corpora (Table~\ref{tab:in_domain}). That accuracy, however, does not reveal the FPR that is actually realized after a transfer to different mail~\citep{quinonero2009shift,ovadia2019trust}.

In this paper, we study this cold-start setting, in which the organization has no labeled legitimate mail of its own, and find that it is where trained detectors fail and where a retrieval design earns its place. Calibrated for a 1\% FPR budget on older public mail and applied to DataPhish 2025, a recent corpus of human- and LLM-written email, word-frequency (TF-IDF) and embedding classifiers realize \NumShiftBaseLo--\NumShiftBaseHi{} FPR on that mail. For a hospital that receives 10{,}000 legitimate emails a day, this would mean roughly 2{,}000 false alarms a day instead of the 100 the budget allows. Even their most conservative calibration, a threshold above every legitimate validation email, leaves \NumShiftZeroFPBest{}. The cause is the shift in \emph{legitimate} mail (\emph{benign-distribution shift}), not the style of the attacks: when benign mail comes from the same distribution, rewriting phishing with LLMs pushes no method's FPR materially above its budget and does not reduce the recall of embedding-based or trained detectors. Nor is the failure one of ranking: given thresholds chosen on the deployment mail itself, the same classifiers separate phishing from legitimate mail well. What breaks is \emph{calibration transfer}: a threshold learned from one corpus's legitimate mail lands in the wrong place on another's. We call the ratio of the FPR realized on deployment mail to the budget the \emph{overrun}.

This diagnosis suggests the remedy. We introduce CyberCane, a detector whose score depends only on deterministic rules and on similarity to \emph{known phishing}. Because no legitimate mail enters the score, the calibration source affects only the threshold, which is why that threshold transfers far better: the overrun falls \NumRtwoFacLo--\NumRtwoFacHi$\times$ (Section~\ref{sec:generality}). One reason to expect this is that phishing tends to look similar across organizations, since a single campaign reaches many inboxes, while legitimate mail differs from one organization to the next. Email text passes a privacy boundary (pattern-based and local named-entity redaction) before any external call; retrieval runs over a training-only, redacted index; and a single risk score is calibrated to an explicit FPR budget. Scoring the most phishing-like segment of an email, rather than the email as a whole, keeps benign padding from diluting that evidence. An ontology encodes attack patterns as rules in the Web Ontology Language (OWL) that are asserted only when every premise holds, and explanations are checked against the stored evidence.

We do not address the case in which an organization already holds labeled legitimate mail of its own. Calibrated on that mail, trained classifiers stay near the budget and reach the highest recall (Table~\ref{tab:in_domain}), so no design question remains. The study answers three research questions. \textit{RQ1:} How far do trained detectors overrun an FPR budget when calibrated on public corpora and deployed on different mail, and what causes the overrun? \textit{RQ2:} Does a phishing-only retrieval rule reduce the overrun, and which component (decision rule or representation) is responsible? \textit{RQ3:} How does this design compare with zero-shot LLMs, which need no training data, and what do the privacy boundary, ontology, and explanation checks contribute?

\textbf{Contributions.} We make the following contributions:

\begin{enumerate}[label=\textbf{(\roman*)}, leftmargin=*, itemsep=1pt, parsep=0pt, topsep=2pt]
\item \textbf{A deployment-risk finding.} Trained phishing detectors calibrated on public corpora overrun a 1\% FPR budget \NumRtwoBaseMinOver--\NumRtwoBaseMaxOver$\times$ under benign-distribution shift. We isolate benign shift, not LLM-written attacks, as the cause, and show that LLM rewriting does not reduce the recall of embedding-based and trained detectors.
\item \textbf{A mechanism and a design principle.} We show that trained detectors fail by calibration transfer, not by ranking, and that scoring by similarity to known phishing removes the dependence of the score on legitimate mail. This phishing-only rule reduces the overrun \NumRtwoFacLo--\NumRtwoFacHi$\times$ across three calibration sources and five seeds, for every tested setting of its score weights, on a second, DataPhish-independent family of shifts, and with both a proprietary and an open, locally run embedder. Scoring the most phishing-like segment lowers the realized FPR a further \NumSegOverRedLo--\NumSegOverRedHi$\times$ at unchanged ranking quality and removes a padding evasion. The rule needs neural embeddings to keep recall on unseen attacks.
\item \textbf{CyberCane.} A privacy-bounded neuro-symbolic system with an FPR-calibrated risk score, standards-conformant sender-authentication rules, ontology rules asserted only on full premise satisfaction, and evidence-checked explanations, with redaction recall measured per identifier type. It screens an email locally in \NumLatLocal{}\,s at no per-email cost and is released with its evaluation harness.
\item \textbf{A benchmark correction.} A random split of DataPhish 2025 places \NumLeakPct{} of test emails in the same rewrite family (an original email and its LLM rewrites) as a training email, and about two-thirds of its phishing test emails belong to rewrite families that near-duplicate the Nazario corpus. We provide a leak-free, group-aware protocol under which all results are reported, with preregistered criteria and cluster-bootstrap intervals.
\end{enumerate}

\FloatBarrier
\section{Related Work}
\label{sec:related}

\textbf{Phishing detection.} Detection evolved from blocklists~\citep{apwg2023q4} to learned classifiers over lexical, URL, and header features~\citep{aleroud2017,shirazi2018}, complemented by sender-authentication protocols~\citep{kitterman2014spf,crocker2011dkim,kucherawy2015dmarc}: SPF lists the servers allowed to send for a domain, DKIM signs messages, and DMARC publishes the domain owner's policy for mail that fails these checks. Such classifiers reach near-ceiling accuracy on in-distribution splits of classic public corpora, which our leak-free experiments reproduce on Nazario and SpamAssassin. Their behavior when the legitimate mail of deployment differs from that of training has received little attention.

\textbf{LLM-based detectors.} LLMs can both write convincing phishing emails and recognize phishing intent in them~\citep{heiding2024devising}. ChatSpamDetector prompts a large language model with the email and asks for a verdict with reasons~\citep{koide2026chatspamdetector}; it needs no training data but sends email content to the model. KnowPhish augments detection with a large brand knowledge base and targets webpage-level, reference-based detection~\citep{li2024knowphish}; its code and knowledge base are not publicly released, which prevents a faithful reproduction on email text. We therefore compare against an open-weight zero-shot LLM with a ChatSpamDetector-style prompt applied to redacted text (Section~\ref{sec:baselines}).

\textbf{Retrieval-augmented and neuro-symbolic security.} Retrieval-augmented generation grounds model outputs in retrieved evidence, and has been applied to cybersecurity question answering~\citep{simoni2025morse} and to neuro-symbolic retrieval with adaptive query routing~\citep{hakim2025symrag}. Neuro-symbolic systems combine learned perception with explicit rules~\citep{garcez2023neurosymbolic,hu2016harnessing}, and their use in cybersecurity is surveyed in~\citep{hakim2025neurosymbolic}. These works use retrieval to improve answers or efficiency; we use retrieval over a phishing-only index as the \emph{decision rule} itself and study its false-positive behavior under shift.

\textbf{Faithful explanations.} LLM-generated explanations can be plausible yet unfaithful: chain-of-thought rationales can misrepresent the actual reason for a prediction, omitting biasing features that influenced it~\citep{turpin2023unfaithful}. Factuality evaluation therefore splits generated text into atomic claims and checks each against a knowledge source~\citep{min2023factscore}. We apply such claim-level checks against the system's own stored evidence rather than an external source, and the verdict never depends on the explanation, so an unfaithful explanation cannot change a decision.

\textbf{False-positive control under shift.} Neyman--Pearson classification bounds the type~I error of any scoring classifier by choosing its threshold on held-out data of the negative class~\citep{tong2018np}, and conformal prediction gives distribution-free guarantees in the same way~\citep{tibshirani2019conformal}. Both guarantees assume that the held-out data are exchangeable with deployment data; weighted conformal prediction relaxes this under covariate shift when the likelihood ratio between the two distributions is known or can be estimated from unlabeled deployment data~\citep{tibshirani2019conformal}. Dataset shift more generally degrades calibrated predictions~\citep{quinonero2009shift,ovadia2019trust} and is hard to detect without labels~\citep{rabanser2019failing}. In phishing, detectors that separate LLM-written from human-written phishing keep an AUROC above 0.96 across generators yet lose 28 F1 points on average at a fixed threshold, a gap that a threshold re-tuned on labeled target emails largely closes~\citep{gutierrez2026crossmodel}; we find the same pattern of intact ranking and failed thresholds for phishing versus legitimate mail, where the deployer has no target labels. Our setting violates exchangeability by construction: the threshold is set on one corpus's legitimate mail and applied to another's. Rather than only correcting the threshold, CyberCane changes what the score depends on, so that legitimate mail enters only the threshold, never the score function; label-free threshold correction is tested separately (Section~\ref{sec:recalibration}).

\textbf{One-class decisions and evaluation pitfalls.} Deciding from similarity to a single class relates to one-class description~\citep{tax2004svdd} and positive--unlabeled learning~\citep{elkan2008pu}; hubness corrections such as cross-domain similarity local scaling (CSLS)~\citep{conneau2018word} adjust nearest-neighbor similarities using unlabeled data, which we test as a label-free adaptation (Section~\ref{sec:negative}). In security, unrepresentative train/test distributions and impossible time splits inflate reported results~\citep{pendlebury2019tesseract}, and data snooping is among the most common pitfalls of learning-based security systems~\citep{arp2022dos}; leakage between training and test data affects many fields~\citep{kapoor2023leakage}. We document such leakage in a current phishing benchmark and evaluate under an explicit shift between the calibration and deployment sources. Table~\ref{tab:related} summarizes how the compared approaches differ.

\begin{table}[!htbp]
\centering
\scriptsize
\renewcommand{\arraystretch}{0.9}
\caption{Mechanism-level comparison of detection approaches relevant to benign-distribution shift and privacy.}
\label{tab:related}
\begin{tabular}{p{0.19\linewidth}p{0.20\linewidth}p{0.16\linewidth}p{0.18\linewidth}p{0.15\linewidth}}
\toprule
Approach & Decision rule & Labeled benign mail needed & Content sent externally & Explanation \\
\midrule
Trained text/embedding classifiers & Learned two-class boundary & Yes (training and calibration) & No (local) or embeddings & Feature weights \\
ChatSpamDetector~\citep{koide2026chatspamdetector} & LLM judgment & No & Email text to the LLM & Generated rationale \\
KnowPhish~\citep{li2024knowphish} & Brand-knowledge reference matching (webpages) & Not applicable (webpages) & Depends on deployment & Brand match \\
RAG for security~\citep{simoni2025morse,hakim2025symrag} & Generation grounded in retrieved text & Task-dependent & Retrieved context & Cited evidence \\
One-class / PU learning~\citep{tax2004svdd,elkan2008pu} & Boundary around one class & No (calibration only) & No & None \\
NP classification / conformal~\citep{tong2018np,tibshirani2019conformal} & Any score; threshold on held-out negatives & Yes (exchangeable with deployment) & No & None \\
\textbf{CyberCane} & Similarity to known phishing + rules & Calibration only & Redacted text, or none (local variant) & Rules, neighbors, checked tags \\
\bottomrule
\end{tabular}
\end{table}

\FloatBarrier
\section{Problem Setting and Threat Model}
\label{sec:setting}

\textbf{Cold-start calibration.} A deploying organization chooses a threshold $t$ on a detector's score $s$ so that the fraction of legitimate emails with $s>t$ does not exceed a budget $b$ (we use $b \in \{0.5\%, 1\%, 2\%\}$). The threshold is estimated on legitimate mail available at adoption time, typically a public corpus, and then applied to the organization's traffic. We call the ratio of the realized FPR on the deployment mail to $b$ the \emph{budget overrun}, and report realized FPR, the overrun, and recall at the realized operating point.

\textbf{Attacker capabilities.} We assume the attacker can (1) register lookalike domains or send from compromised domains that pass SPF/DKIM; (2) host links on legitimate infrastructure (URL shorteners, file-sharing services) that evades blocklists; (3) generate fluent, domain-specific content with LLMs~\citep{dipalma2025airisks}; and (4) exploit domain workflows and pressure points, such as appointment, prescription, and insurance themes in healthcare.

\textbf{Scope.} In scope are text-bearing emails with embedded links, urgency-based social engineering, credential harvesting, and impersonation. Out of scope are image-only and QR-code phishing, malicious attachments, voice/SMS channels, and fully authenticated insider mail from compromised accounts.

\textbf{Privacy threat model.} The defender runs the detector and may call external embedding or language-model services. Anything sent to such a service is treated as disclosed. The privacy goal is therefore to minimize identifying content that crosses this boundary, and to measure what remains; we do not claim formal de-identification under HIPAA's Safe Harbor or Expert Determination routes~\citep{hhs2012deid}. Embeddings of redacted text can still leak information under inversion attacks~\citep{li2023sentence,carlini2021extracting}, which motivates the local-embedding configuration discussed in Section~\ref{sec:discussion}.

\FloatBarrier
\section{CyberCane}
\label{sec:method}

\begin{figure}[!htbp]
    \centering
    \includegraphics[width=\textwidth]{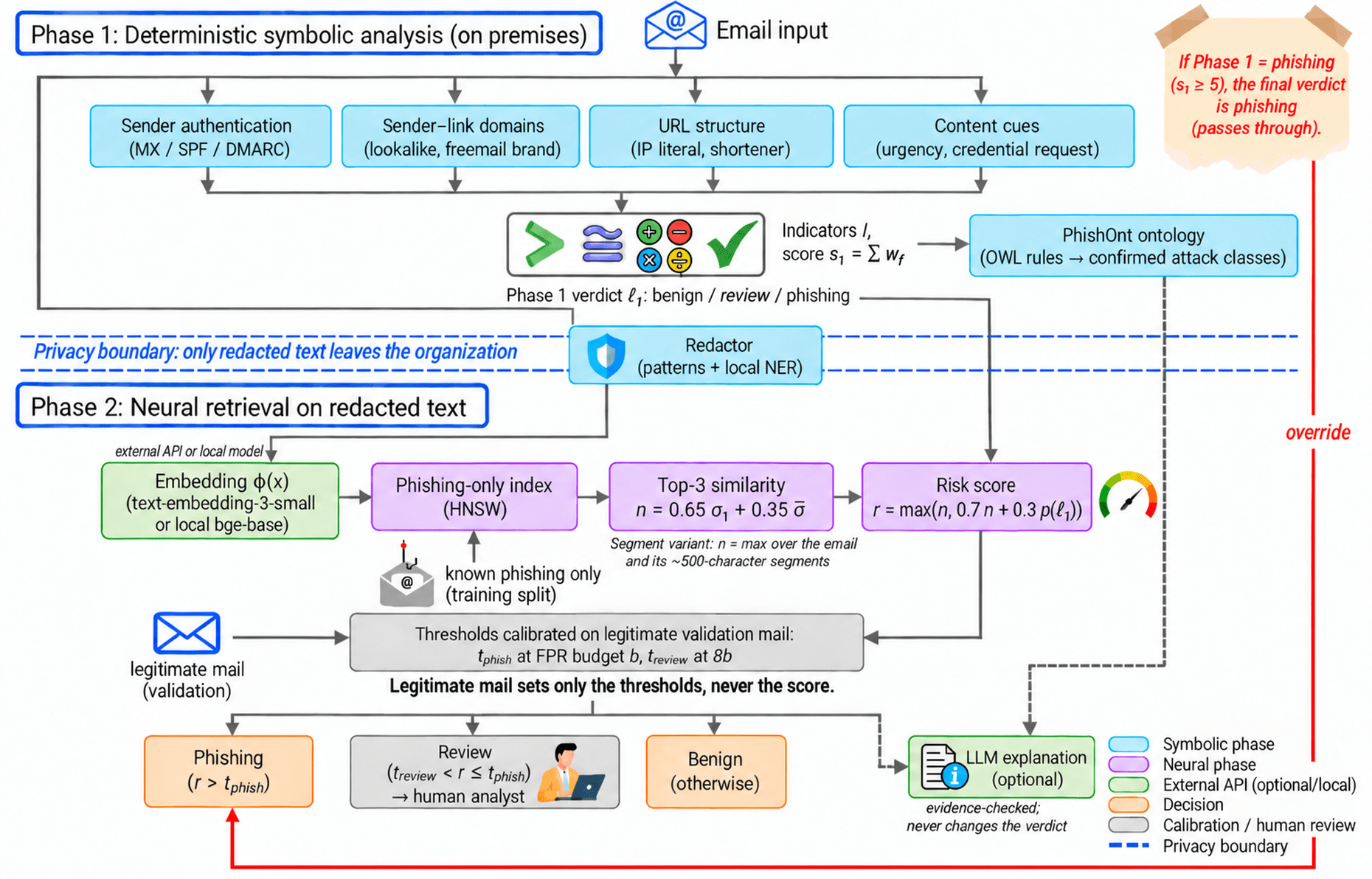}
\caption{CyberCane pipeline. Phase~1 (cyan) analyzes the raw email on premises: weighted rules yield indicators $I$, a score $s_1$, and a verdict $\ell_1$, and PhishOnt asserts attack classes from the indicators; a Phase~1 phishing verdict ($s_1 \geq 5$) passes through unchanged. The redactor is the gate of the privacy boundary (dashed lines), so Phase~2 (lavender) sees only redacted text. Retrieval uses an index of training phishing only (CyberCane-Seg also scores segments of about 500 characters; Section~\ref{sec:retrieval}), and the risk score $r$ combines the similarity score $n$ with the Phase~1 prior $p(\ell_1)$ (Eq.~\ref{eq:risk}). The thresholds $t_{\text{phish}}$ and $t_{\text{review}}$ are calibrated on legitimate mail for an FPR budget $b$ and a review budget $8b$, so legitimate mail sets only the thresholds, never the risk score. Green components can run as an external API or locally. The optional explanation is checked against the Phase~1 and PhishOnt evidence and never changes the verdict.}
    \label{fig:architecture}
\end{figure}

CyberCane processes an email in two phases (Fig.~\ref{fig:architecture}). Phase~1 is deterministic and local; Phase~2 is neural and calls external services only with redacted text. The verdict is produced by a calibrated risk score, not by a language model, so it is reproducible and auditable.

\subsection{Phase 1: Deterministic Symbolic Analysis}
\label{sec:phase1}

Phase~1 applies weighted rules over five signal families: sender authentication (the domain's mail-exchanger (MX), SPF, and DMARC records~\citep{kitterman2014spf,kucherawy2015dmarc}), sender--link domain relations (lookalike domains, freemail senders claiming a brand), URL structure (IP literals, shorteners), urgency language, and credential requests. Weights $w_f \in \{1,2,3\}$ for each fired rule $f$ are summed into a score $s_1$ that maps to a verdict $\ell_1$: benign ($s_1<2$), needs\_review ($2 \leq s_1 < 5$), or phishing ($s_1 \geq 5$) (Algorithm~\ref{alg:symbolic}; rules in Appendix~\ref{app:rule-details}). Authentication checks follow the email standards. A domain without MX records but with an address record still receives mail (implicit MX)~\citep{klensin2008smtp}; a subdomain without a DMARC record inherits the policy of its organizational domain~\citep{kucherawy2015dmarc}; and a lookup that fails is treated as unknown rather than as a missing record. Phase~1 provides instant, independently checkable evidence and a hard override when several strong rules fire together. On the header-less public corpora used here, DNS checks are disabled for historical domains, and Phase~1 acts mainly as triage: it routes emails to review and contributes a prior to the risk score.

\begin{algorithm2e}[H]
\caption{Phase 1: Deterministic Symbolic Analysis}
\label{alg:symbolic}
\footnotesize
\linespread{0.85}\selectfont
\DontPrintSemicolon
\SetKwProg{Fn}{Function}{}{end}
\KwIn{Email $e$ with headers $H$, body $B$, URLs $U$, rule weights $w_f$}
\KwOut{Indicators $I$, score $s_1$, verdict $\ell_1$}
\Fn{DeterministicAnalysis($e$)}{
    $d \leftarrow \text{domain}(H.\text{from})$;\quad $I \leftarrow \emptyset$\;
    $I \leftarrow I \cup \text{CheckDNS}(d) \cup \text{CheckAuth}(d)$ \tcp{failed lookups count as unknown}
    $I \leftarrow I \cup \text{DomainRelations}(U, d) \cup \text{AnalyzeURLs}(U) \cup \text{ContentCues}(B)$\;
    $s_1 \leftarrow \sum_{f \in I} w_f$\;
    $\ell_1 \leftarrow$ benign if $s_1<2$; needs\_review if $s_1<5$; phishing otherwise\;
    \Return{$(I, s_1, \ell_1)$}
}
\end{algorithm2e}

\subsection{Privacy Boundary}
\label{sec:privacy}

Every piece of email-derived text sent to an external service (query embeddings, index construction, explanation prompts, and the LLM baseline) passes a single function, so the boundary can be audited by inspecting its call sites. The redactor masks identifiers completely rather than preserving digit tails. It combines patterns for email local parts, phone numbers, SSNs, card numbers, dates, labeled record and account identifiers (e.g., MRN, member ID), ZIP codes, street addresses, and digit runs of six or more, with a local BERT-based named-entity recognizer~\citep{devlin2019bert,tjong2003conll} for person names and locations. Phishing signals are deliberately kept: URL hosts and paths, sender and link domains, brand names, and a marker for userinfo tricks such as \texttt{paypal.com@evil.example}. URL query strings are dropped. The same sanitized representation is used for index construction and for queries, so both sides of a similarity comparison share one text space.

\subsection{Phishing-Only Retrieval and Risk Score}
\label{sec:retrieval}

CyberCane asks one question of an email: how closely does it resemble attacks seen before? The redacted text $x$ is embedded as $\phi(x)$ (text-embedding-3-small, 1536 dimensions, unit norm) and compared by cosine similarity with a phishing-only index $\mathcal{P}$ built from the training split. Let $\sigma_1 \geq \sigma_2 \geq \sigma_3$ be the three largest similarities and $\bar{\sigma} = (\sigma_1+\sigma_2+\sigma_3)/3$ their mean. The neighbor signal and the risk score are
\begin{equation}
n(x) = 0.65\,\sigma_1 + 0.35\,\bar{\sigma}, \qquad
r(x) = \operatorname{clip}_{[0,1]}\!\big(\max\{\,n(x),\; 0.7\,n(x) + 0.3\,p(\ell_1)\,\}\big),
\label{eq:risk}
\end{equation}
where $p(\text{phishing})=1$, $p(\text{needs\_review})=0.6$, and $p(\text{benign})=0$. The maximum ensures that Phase~1 can raise but never lower the semantic evidence. The risk score is a ranking score, not a probability; it is displayed as $\operatorname{round}(10\,r)$. Because no legitimate mail enters $\mathcal{P}$, the score measures resemblance to known attacks rather than distance from one corpus's notion of normal mail. In particular, the score of an email does not change when a different legitimate corpus is used for calibration; only the threshold does. This is the property our experiments test.

\textbf{Segment-level scoring.} A whole-email embedding averages over all of its text, so an attacker can dilute it by appending benign text. CyberCane-Seg also splits the redacted text into consecutive segments $x_1, \dots, x_m$ of about 500 characters, computes $n$ for each, and uses the largest value, $n_{\text{seg}}(x) = \max\{n(x), \max_j n(x_j)\}$, in place of $n(x)$ in Eq.~(\ref{eq:risk}). The index is unchanged, and the calibration procedure is the same.

\subsection{Calibrated Decision Rule}
\label{sec:decision}

Two thresholds are calibrated on validation legitimate mail: $t_{\text{phish}}$ at the FPR budget $b$ and $t_{\text{review}}$ at a wider review budget ($8b$). Put simply, at a 1\% budget an email is flagged when its risk is higher than that of 99\% of the legitimate validation emails. Algorithm~\ref{alg:decision} gives the verdict. A Phase~1 phishing verdict passes through unchanged; on the public corpora this path fires for at most \NumPhaseOneOverride{} of emails, so the evaluated rule is essentially the calibrated risk threshold.

\begin{algorithm2e}[H]
\caption{Phase 2: Calibrated Retrieval Decision}
\label{alg:decision}
\footnotesize
\linespread{0.85}\selectfont
\DontPrintSemicolon
\SetKwProg{Fn}{Function}{}{end}
\KwIn{Email $e$, Phase 1 $(I, \ell_1)$, phishing index $\mathcal{P}$, thresholds $t_{\text{phish}}$, $t_{\text{review}}$}
\KwOut{Verdict $\ell$, risk $r$, evidence $E$}
\Fn{CalibratedDecision($e, I, \ell_1$)}{
    $x \leftarrow \text{PrivacyBoundary}(e)$ \tcp{redact before any external call}
    $N \leftarrow \text{TopK}\big(\cos(\phi(x), \mathcal{P}), k{=}8\big)$ \tcp{training-only, phishing-only index}
    $r \leftarrow$ Eq.~\eqref{eq:risk} on $N$ and $\ell_1$\;
    $\ell \leftarrow$ phishing if $\ell_1{=}\text{phishing} \lor r>t_{\text{phish}}$; needs\_review if $r>t_{\text{review}}$; benign otherwise\;
    $E \leftarrow (I, N, r, \text{OntologyMatch}(I))$ \tcp{stored for explanation checking}
    \Return{$(\ell, r, E)$}
}
\end{algorithm2e}

\subsection{Ontology-Informed Rule Matching}
\label{sec:ontology}

PhishOnt, CyberCane's phishing ontology, encodes attack patterns over Phase~1 indicators as OWL classes defined by \texttt{owl:equivalentClass} axioms (Table~\ref{tab:ontology_rules}). The rules executed at run time are parsed from these axioms, so the ontology is their single source of truth; a lightweight matcher evaluates them rather than a description-logic reasoner. Algorithm~\ref{alg:ontology} separates three outcomes. A class is \emph{asserted} only when every premise holds; its superclasses are \emph{inherited} through \texttt{rdfs:subClassOf}; and a class with at least half of its premises satisfied is reported as a \emph{suggestion} together with the missing premises, but never asserted. The coverage ratio of a suggestion is a matching statistic, not a probability. Two parent-level patterns are modeled as defined subclasses (\texttt{MissingMailAuthentication}, \texttt{UrgentCredentialRequest}), because an equivalence axiom on a parent class would force every subclass to satisfy it. The ontology contributes structured, checkable evidence for explanations; it does not change the verdict.

\begin{table}[!htbp]
\centering
\scriptsize
\renewcommand{\arraystretch}{0.9}
\caption{PhishOnt attack classes and their defining premises, parsed from the ontology file.}
\label{tab:ontology_rules}
\begin{tabular}{p{0.30\linewidth}p{0.42\linewidth}p{0.20\linewidth}}
\toprule
Defined class & Premises (\texttt{owl:equivalentClass}) & Inherited superclasses \\
\midrule
\texttt{AppointmentScam} & UrgencyLanguage $\wedge$ FreemailSender & SocialEngineeringAttack \\
\texttt{CredentialTheft} & CredentialRequest $\wedge$ MissingMX & TechnicalAttack \\
\texttt{HighConfidencePhishing} & UrgencyLanguage $\wedge$ CredentialRequest $\wedge$ MissingDMARC & --- \\
\texttt{InsuranceVerificationPhish} & CredentialRequest $\wedge$ FreemailSender & SocialEngineeringAttack \\
\texttt{MissingMailAuthentication} & MissingMX $\wedge$ MissingSPF & TechnicalAttack \\
\texttt{PrescriptionFraud} & UrgencyLanguage $\wedge$ DomainMismatch & SocialEngineeringAttack \\
\texttt{URLBasedAttack} & DomainMismatch $\wedge$ (IPLiteralURL $\vee$ ShortenedURL) & TechnicalAttack \\
\texttt{UrgentCredentialRequest} & UrgencyLanguage $\wedge$ CredentialRequest & SocialEngineeringAttack \\
\bottomrule
\end{tabular}

\end{table}

\begin{algorithm2e}[H]
\caption{Ontology-Informed Rule Matching}
\label{alg:ontology}
\footnotesize
\linespread{0.85}\selectfont
\DontPrintSemicolon
\SetKwProg{Fn}{Function}{}{end}
\KwIn{Phase 1 indicators $I$; for each class $\tau$, its all-of premises $\mathcal{R}^{\wedge}_\tau$ and any-of premises $\mathcal{R}^{\vee}_\tau$, parsed from the ontology}
\KwOut{Asserted classes $\mathcal{A}$, inherited classes $\mathcal{A}^{+}$, suggestions $\mathcal{S}$}
\Fn{OntologyMatch($I$)}{
    $\Pi \leftarrow \{\text{MapProperty}(i) : I[i]=\text{true}\}$ \tcp{premises that hold}
    \For{each defined class $\tau$}{
        $M \leftarrow \mathcal{R}^{\wedge}_\tau \setminus \Pi$, plus ``one of $\mathcal{R}^{\vee}_\tau$'' if $\mathcal{R}^{\vee}_\tau \neq \emptyset$ and $\mathcal{R}^{\vee}_\tau \cap \Pi = \emptyset$ \tcp{missing premises}
        $c \leftarrow 1 - |M| / \big(|\mathcal{R}^{\wedge}_\tau| + \mathbbm{1}[\mathcal{R}^{\vee}_\tau \neq \emptyset]\big)$\;
        \lIf{$M=\emptyset$}{$\mathcal{A} \leftarrow \mathcal{A} \cup \{\tau\}$}
        \lElseIf{$c \geq 0.5$}{$\mathcal{S} \leftarrow \mathcal{S} \cup \{(\tau, c, M)\}$}
    }
    $\mathcal{A}^{+} \leftarrow \text{Superclasses}(\mathcal{A}) \setminus \mathcal{A}$\;
    \Return{$(\mathcal{A}, \mathcal{A}^{+}, \mathcal{S})$}
}
\end{algorithm2e}

\subsection{Evidence-Checked Explanations}
\label{sec:explanations}

An LLM receives the redacted email, the Phase~1 indicators, the retrieved neighbors, and the asserted ontology classes, and writes short tagged explanations ([AUTH], [URL], [URGENCY], [CONTENT], [SIMILARITY], [ONTOLOGY]). The verdict never depends on this text. Each tagged claim can be checked against the stored evidence $E$ (Section~\ref{sec:system}). Four checks apply: cited similarities must match a retrieved neighbor or a summary of the retrieved similarities within 0.01, and none may be cited when nothing was retrieved; named attack classes must be asserted, not merely suggested; authentication and URL claims must agree with the recorded indicators; quoted phrases must occur in the email. A claim is \emph{supported}, \emph{contradicted}, or \emph{unverifiable}, and in evaluation every generated claim stays in the denominator. Fig.~\ref{fig:explanation_example} (Section~\ref{sec:system}) shows a checked explanation.

\FloatBarrier
\section{Experimental Protocol}
\label{sec:protocol}

\subsection{Corpora and Leak-Free Splitting}
\label{sec:data}

We use Nazario phishing~\citep{nazario2007phishing} with SpamAssassin~\citep{spamassassin2006corpus}; DataPhish 2025~\citep{toth2025dataphish}, in which each of \NumDPFamilies{} original emails appears in four versions (one human-written and three LLM rewrites); and, as additional legitimate-mail (``ham'') sources, CEAS 2008 ham (a random sample of 4{,}000) and Ling-Spam ham~\citep{androutsopoulos2000lingspam}, taken from a compiled public collection~\citep{alsubaiey2024phishing}. Aggregated collections that re-package these corpora are excluded to avoid cross-corpus duplication. We use the DataPhish release of the preprint's version~2 (January 2026); later versions describe a different, regenerated dataset.

This matters because a test email with a near-identical copy in training could be scored correctly by matching the copy rather than by recognizing phishing. Splits are group-aware. Exact duplicates (after normalizing case, whitespace, and URLs to hosts), near-duplicates (character 3--5-gram TF-IDF cosine $\geq 0.90$), and DataPhish rewrite families are merged into groups with union--find, and groups are split 70/10/20 with stratification for five seeds. No group, exact text, or rewrite family is shared between splits of any seed. DataPhish reduces to \NumDPGroups{} independent groups (Table~\ref{tab:splits}). For every cross-corpus evaluation, test groups with any member at character TF-IDF cosine $\geq 0.80$ to the source training split are removed (\emph{decontamination}).

\begin{table}[!htbp]
\centering
\scriptsize
\renewcommand{\arraystretch}{0.9}
\caption{Corpora after leak-free grouping (seed 42). P/S/L: phishing/spam/legitimate.}
\label{tab:splits}
\begin{tabular}{lrrccc}
\toprule
Corpus & Rows & Groups & Train (P/S/L) & Val (P/S/L) & Test (P/S/L) \\
\midrule
Nazario + SpamAssassin & 7,374 & 6,638 & 1,096 / 1,203 / 2,864 & 157 / 171 / 409 & 312 / 344 / 818 \\
DataPhish 2025 & 12,300 & 2,699 & 3,152 / 2,800 / 2,660 & 448 / 400 / 380 & 900 / 800 / 760 \\
CEAS 2008 ham (sample) & 4,000 & 3,813 & 0 / 0 / 2,800 & 0 / 0 / 400 & 0 / 0 / 800 \\
Ling-Spam ham & 2,401 & 2,270 & 0 / 0 / 1,681 & 0 / 0 / 240 & 0 / 0 / 480 \\
\bottomrule
\end{tabular}

\end{table}

\subsection{Tasks and Shift Settings}
\label{sec:settings}

The task is phishing versus legitimate mail; spam is excluded from evaluation, because SpamAssassin spam is not phishing (a phishing-or-spam task is summarized in Section~\ref{sec:negative}). We evaluate three settings:
\begin{itemize}[leftmargin=*, itemsep=1pt, parsep=0pt, topsep=2pt]
\item \textbf{Benign-distribution shift (primary).} Train and calibrate on Nazario phishing plus an older legitimate-mail source (SpamAssassin, CEAS, or Ling-Spam); test on decontaminated DataPhish 2025.
\item \textbf{Reverse shift.} Train and calibrate on DataPhish; test on Nazario phishing plus each older legitimate source.
\item \textbf{Attack-style shift.} Within DataPhish, train and calibrate on human-written emails and test on LLM-written emails of disjoint families, with a human-to-human control. Legitimate mail comes from the same distribution on both sides. Because the human-written validation split is small, thresholds come from out-of-fold scores of a five-fold group-wise cross-validation on the source.
\end{itemize}

\subsection{Baselines}
\label{sec:baselines}

All methods receive the same redacted text. \textbf{B3}: TF-IDF (unigrams and bigrams) with logistic regression. \textbf{B9}: logistic regression on the same embeddings as CyberCane. \textbf{B10}: a two-class $k$-nearest-neighbor score (mean top-$k$ similarity to phishing minus that to legitimate mail). \textbf{B11}: CyberCane's phishing-only scoring rule with TF-IDF vectors instead of embeddings. \textbf{B12}: a zero-shot LLM, the open-weight \texttt{gpt-oss-120b}~\citep{openai2025gptoss}, prompted in the style of ChatSpamDetector to return a phishing probability (Appendix~\ref{app:prompts}), pinned to a single bf16 serving provider for reproducibility (\NumLLMCalls{} calls, \NumLLMFailures{} failures). B9--B11 isolate the two components of CyberCane: B9 and B10 share its representation but not its decision rule, and B11 shares its decision rule but not its representation. KnowPhish is not reproduced because its code and knowledge base are not released.

\subsection{Calibration, Metrics, and Statistics}
\label{sec:metrics}

Every method outputs a continuous score. Thresholds are set only on source-validation legitimate mail under three rules: the budget quantile, a safety rule (budget$/4$), and a zero-false-positive rule (the maximum validation score). The two stricter rules stand in for a cautious operator who sets the threshold tighter than the budget, and they show whether such caution is enough on its own. No target labels are used for calibration. We report realized target FPR and recall, and, where ranking quality matters, the threshold-free partial AUROC (pAUC): the area under the ROC curve restricted to FPR $\leq 2\%$, McClish-normalized so that 0.5 is chance, which summarizes how well a score ranks phishing above legitimate mail at the strict end. Results are means over five split seeds. Because DataPhish rewrites of one original are correlated, FPR confidence intervals use a cluster bootstrap over groups (1{,}000 resamples)~\citep{field2007cluster}; they are about twice as wide as Clopper--Pearson intervals~\citep{clopper1934}. Each experiment, including its success criterion and the interpretation of each outcome, was recorded in a dated log before it was run (Appendix~\ref{app:prereg}).

\subsection{Implementation}
\label{sec:implementation}

CyberCane is implemented as a FastAPI service (Python 3.11) with PostgreSQL and pgvector cosine HNSW indexes~\citep{malkov2020hnsw} and a Next.js review interface. Retrieval restricts candidates to training-split phishing rows. After index construction, an indexed lookup takes about 3.5\,ms, and the service's retrieval matches exact search (mean top-8 overlap 0.998 over 50 test queries). Experiments use cached embeddings, and all scoring runs on CPU. The zero-shot baseline cost \NumLLMCostTotal{} in total, \NumLLMCostPerK{} per 1{,}000 emails.

\FloatBarrier
\section{Results}
\label{sec:results}

We first check the benchmark itself, then measure false-positive budgets under shift and their generality (RQ1), isolate what drives the effect and how segment-level scoring and recalibration change it (RQ2), and close with zero-shot LLMs and the evidence for the privacy boundary, authentication rules, ontology, and explanations (RQ3).

\subsection{Benchmark Integrity}
\label{sec:leakage}

Table~\ref{tab:leakage} audits the random-split DataPhish 2025 evaluation used in an earlier version of this work. With a random split, \NumLeakRows{} of \NumLeakTotal{} test emails (\NumLeakPct{}) have a rewrite of the same original in the training split. A retrieval detector then largely recognizes a sibling: RAG recall is \NumOldRecall{} overall but markedly lower on the few test emails without a sibling. The FPR on this split (\NumOldFPR{}) was also far above the low rate obtained on SpamAssassin, which a combined headline obscured. In addition, about \NumNazOverlapPos{} of DataPhish's phishing test emails per seed belong to rewrite families with a member that near-duplicates a Nazario training email, because DataPhish's human-written phishing overlaps that corpus. All results below use the group-aware, decontaminated protocol of Section~\ref{sec:data}.

\begin{table}[!htbp]
\centering
\scriptsize
\renewcommand{\arraystretch}{0.9}
\caption{Leakage audit of the DataPhish 2025 random split and of cross-corpus overlap.}
\label{tab:leakage}
\begin{tabular}{lr}
\toprule
Audit item & Value \\
\midrule
DataPhish test rows with a rewrite sibling in training (old split) & 2,211 / 2,300 (96.1\%) \\
Old RAG recall on DataPhish, all test rows & 99.1\% \\
\quad on leaked rows & 99.7\% \\
\quad on rows without a sibling in training (n=89) & 82.1\% \\
Old RAG FPR on DataPhish (reported as ``---'') & 4.1\% \\
DataPhish phishing test rows near-duplicating Nazario/SpamAssassin training (v2, mean of 5 seeds) & 602 of 900 \\
DataPhish rows / independent groups after leak-free grouping & 12,300 / 2,699 \\
\bottomrule
\end{tabular}

\end{table}

\subsection{False-Positive Budgets Under Benign-Distribution Shift}
\label{sec:shift}

Table~\ref{tab:main_shift} reports the primary setting: calibration on Nazario and SpamAssassin, deployment on DataPhish 2025. Every trained baseline overruns a 1\% budget by a factor of about twenty, with intervals far from the budget, whether it uses TF-IDF or the same embeddings as CyberCane. CyberCane realizes \NumShiftCCFPR{} at \NumShiftCCRecall{} recall, and its FPR varies much less across seeds. The overrun is not a thresholding artifact: under the most conservative rule, which admits no false positive on validation, the best trained baseline still realizes \NumShiftZeroFPBest{}, whereas CyberCane realizes \NumShiftCCZeroFP{} at \NumShiftCCZeroFPRecall{} recall. Within a single corpus, by contrast, all methods stay near the budget and the embedding-based baselines reach the highest recall (Table~\ref{tab:in_domain}), so the advantage is specific to the shifted deployment that motivates this work.

\begin{table}[!htbp]
\centering
\scriptsize
\renewcommand{\arraystretch}{0.9}
\caption{Calibration on Nazario + SpamAssassin, deployment on decontaminated DataPhish 2025 (phishing vs.\ legitimate; mean$\pm$SD over five seeds; 95\% cluster-bootstrap CI of FPR at the 1\% budget).}
\label{tab:main_shift}
\begin{tabular}{lccccc}
\toprule
 & \multicolumn{3}{c}{1\% FPR budget} & \multicolumn{2}{c}{Zero-FP calibration} \\
\cmidrule(lr){2-4}\cmidrule(lr){5-6}
Method & FPR (\%) & 95\% CI & Recall & FPR (\%) & Recall \\
\midrule
TF-IDF LR (B3) & 22.9$\pm$9.4 & [17.4, 29.1] & 98.2\% & 13.6$\pm$6.2 & 94.3\% \\
Embedding LR (B9) & 20.2$\pm$4.2 & [14.8, 26.3] & 99.7\% & 8.9$\pm$2.2 & 97.4\% \\
Two-class $k$NN (B10) & 22.5$\pm$3.6 & [16.6, 28.7] & 100.0\% & 10.9$\pm$3.9 & 97.1\% \\
Phishing-only TF-IDF $k$NN (B11) & 4.5$\pm$3.2 & [2.2, 7.1] & 79.8\% & 0.1$\pm$0.2 & 25.8\% \\
CyberCane, retrieval only & 6.0$\pm$2.1 & [2.9, 9.6] & 87.8\% & 2.4$\pm$1.9 & 83.0\% \\
\textbf{CyberCane} & 5.4$\pm$1.8 & [2.4, 8.8] & 86.4\% & 1.2$\pm$0.6 & 77.3\% \\
\midrule
Zero-shot LLM (B12) & 0.7$\pm$0.6 & [0.1, 1.7] & 56.4\% & 0.2$\pm$0.1 & 22.5\% \\
\bottomrule
\end{tabular}

\end{table}

\subsection{Generality Across Calibration Sources}
\label{sec:generality}

\begin{figure}[!htbp]
    \centering
    \begin{subfigure}[t]{264.7bp}
        \includegraphics{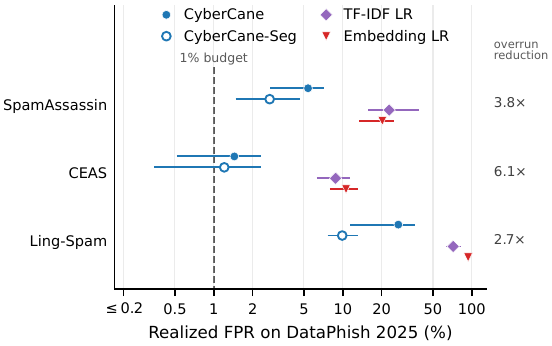}
        \caption{Realized FPR on DataPhish 2025 at a 1\% budget set on each older source (mean of five seeds; bars: seed range; seeds with no false positive at the axis floor). Open circles: CyberCane-Seg, at a cost of \NumSegRecCostLo--\NumSegRecCostHi{} recall points (Table~\ref{tab:segment}). Right: whole-email CyberCane's overrun reduction against the better trained baseline.}
        \label{fig:fpr_budget_a}
    \end{subfigure}\hfill
    \begin{subfigure}[t]{222bp}
        \includegraphics{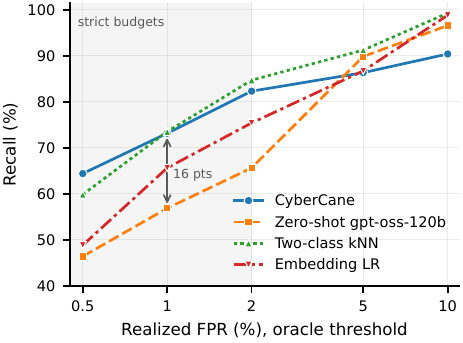}
        \caption{Recall at fixed realized FPR with thresholds chosen on the deployment mail (oracle, not deployable; SpamAssassin calibration, cloud embedder, mean of five seeds).}
        \label{fig:fpr_budget_b}
    \end{subfigure}
    \caption{Calibration fails, ranking does not. Trained baselines rank the deployment mail well (b), so their overrun in (a) is a failure of calibration transfer. CyberCane reduces the overrun under every source and leads at strict operating points, where the zero-shot gpt-oss LLM trails by \NumFrGapOne{} points at 1\% FPR; that LLM overtakes at lenient ones.}
    \label{fig:fpr_budget}
\end{figure}

Fig.~\ref{fig:fpr_budget}a and Table~\ref{tab:benign_sources} repeat the experiment with two further legitimate-mail sources. The overrun of trained baselines ranges from \NumRtwoBaseMinOver$\times$ to \NumRtwoBaseMaxOver$\times$ the budget depending on the source. CyberCane reduces it \NumRtwoFacLo--\NumRtwoFacHi$\times$ on average, and its FPR is below both trained baselines in every seed of every source (preregistered criterion met for 3/3 sources). The absolute FPR, however, depends on how representative the calibration mail is: with broad mailing-list traffic (CEAS) CyberCane stays close to the budget (\NumRtwoCeasCC{}), whereas with a single narrow academic list (Ling-Spam) every method fails, CyberCane included (\NumRtwoLingCC{}). Calibration-set representativeness is therefore a deployment requirement, not a detail. In the reverse direction, calibrating on modern DataPhish mail and deploying on older corpora, no method overruns its budget on average, although single seeds reach up to \NumRevSeedMax{}.

\textbf{A second, DataPhish-independent family of shifts.} To rule out an effect specific to DataPhish, we keep the phishing fixed (Nazario, disjoint groups) and shift only the legitimate mail among the three older sources (Table~\ref{tab:b2_generality}). The trained baselines overrun the budget in \NumBtwoOverPairs{} of 6 pairs (\NumBtwoBaseLo--\NumBtwoBaseHi{} FPR), and in every one of them CyberCane has lower FPR than both trained baselines in all five seeds (\NumBtwoCCLo--\NumBtwoCCHi{}). Calibrating on the broad CEAS traffic keeps the logistic-regression baselines and CyberCane at or near the budget, again pointing to the representativeness of the calibration set.

\textbf{A second, locally run embedder.} The results above use a cloud embedding model. We repeated the main shift with an open model run locally (bge-base-en-v1.5~\citep{xiao2024cpack}), chosen before evaluation and without further model selection, leaving scorers, weights, and thresholds unchanged (Table~\ref{tab:c6_local}). The calibration advantage replicates: CyberCane has lower FPR than both trained baselines in all five seeds of every source, \NumCsixFacLo--\NumCsixFacHi$\times$ lower than the better of them, whereas the two-class $k$NN, which now ranks best, realizes \NumCsixBTenLo--\NumCsixBTenHi{} FPR. The effect therefore follows from how the score is formed rather than from one particular embedding model. Ranking quality does depend on the model: with the local embedder CyberCane loses \NumCsixRecLossLo--\NumCsixRecLossHi{} recall points at the 1\% budget.

\begin{table}[!htbp]
\centering
\scriptsize
\renewcommand{\arraystretch}{0.9}
\caption{Legitimate-mail-only shift: phishing fixed (Nazario), legitimate mail shifted among the older sources (FPR in \%, 1\% budget, mean of five seeds). Overrun: better trained baseline above 2\%. The label-free columns are discussed in Section~\ref{sec:recalibration}.}
\label{tab:b2_generality}
\begin{tabular}{lcccccc}
\toprule
 & \multicolumn{4}{c}{Source-calibrated (1\% budget)} & \multicolumn{2}{c}{Label-free mixture} \\
\cmidrule(lr){2-5}\cmidrule(lr){6-7}
Calibration $\rightarrow$ deployment legitimate mail & Better baseline FPR & CyberCane FPR & CC better (seeds) & Overrun & Best baseline FPR & CyberCane FPR \\
\midrule
CEAS $\rightarrow$ Ling-Spam & 0.3 & \textbf{0.0} & 3/5 & no & 0.2 & 1.8 \\
CEAS $\rightarrow$ SpamAssassin & 0.1 & \textbf{0.1} & 1/5 & no & 1.3 & 1.7 \\
Ling-Spam $\rightarrow$ CEAS & 58.4 & \textbf{16.0} & 5/5 & yes & 1.1 & 3.9 \\
Ling-Spam $\rightarrow$ SpamAssassin & 55.5 & \textbf{11.0} & 5/5 & yes & 0.9 & 1.7 \\
SpamAssassin $\rightarrow$ CEAS & 11.1 & \textbf{3.5} & 5/5 & yes & 3.4 & 3.9 \\
SpamAssassin $\rightarrow$ Ling-Spam & 7.6 & \textbf{0.0} & 5/5 & yes & 0.9 & 1.8 \\
\bottomrule
\end{tabular}

\end{table}

\begin{table}[!htbp]
\centering
\scriptsize
\renewcommand{\arraystretch}{0.9}
\caption{Cloud versus local embedder under the main shift (calibration on the older source, deployment on DataPhish 2025; 1\% budget, mean of five seeds). B3 uses no embeddings. pAUC is McClish-normalized.}
\label{tab:c6_local}
\begin{tabular}{lcccccc}
\toprule
 & \multicolumn{2}{c}{CyberCane FPR / recall (\%)} & \multicolumn{2}{c}{Local embedder, 1\% budget} & \multicolumn{2}{c}{Local pAUC$_{\leq 2\%}$} \\
\cmidrule(lr){2-3}\cmidrule(lr){4-5}\cmidrule(lr){6-7}
Calibration source & Cloud & Local & Better of B3/B9 FPR & Two-class $k$NN FPR / recall & CyberCane & Two-class $k$NN \\
\midrule
SpamAssassin & 5.4 / 86.4 & \textbf{0.6 / 60.8} & 12.4 & 16.6 / 99.3 & 0.812 & 0.847 \\
CEAS & 1.4 / 78.4 & \textbf{1.3 / 63.6} & 7.2 & 8.1 / 95.7 & 0.810 & 0.858 \\
Ling-Spam & 26.9 / 94.7 & \textbf{12.8 / 83.4} & 71.7 & 92.3 / 99.9 & 0.812 & 0.841 \\
\bottomrule
\end{tabular}

\end{table}

\begin{table}[!htbp]
\centering
\scriptsize
\renewcommand{\arraystretch}{0.9}
\caption{Benign-distribution shift across calibration sources (FPR / recall at a 1\% budget, mean of five seeds). Overrun reduction: FPR of the better trained baseline divided by CyberCane's FPR. Reverse: largest seed-mean FPR of any method when calibrating on DataPhish and deploying on the older source.}
\label{tab:benign_sources}
\begin{tabular}{lcccccc}
\toprule
 & \multicolumn{4}{c}{FPR (\%) / recall (\%) at a 1\% budget} & Overrun & Reverse \\
\cmidrule(lr){2-5}
Calibration benign source & TF-IDF LR & Emb.\ LR & Phish-only TF-IDF & CyberCane & reduction & max FPR (\%) \\
\midrule
SpamAssassin (2002) & 22.9 / 98.2 & 20.2 / 99.7 & 4.5 / 79.8 & \textbf{5.4 / 86.4} & 3.8$\times$ & 0.2 \\
CEAS (2008) & 8.8 / 93.7 & 10.6 / 98.9 & 2.6 / 73.2 & \textbf{1.4 / 78.4} & 6.1$\times$ & 0.9 \\
Ling-Spam (list archive) & 71.7 / 99.6 & 93.6 / 100.0 & 19.5 / 89.2 & \textbf{26.9 / 94.7} & 2.7$\times$ & 0.3 \\
\bottomrule
\end{tabular}

\end{table}

\subsection{What Drives the Effect}
\label{sec:mechanism}

Table~\ref{tab:mechanism} crosses representation with decision rule. In other words, one ingredient changes at a time: how the text is represented (word counts or embeddings) and how the decision is made (two-class or phishing-only). Both two-class variants overrun the budget about twentyfold, whether they use TF-IDF or embeddings; both phishing-only variants reduce the overrun about fourfold. The decision rule, not the representation, is therefore what limits false positives. The representation determines recall: at the same budget and a similar realized FPR, CyberCane recovers more phishing than its TF-IDF counterpart (\NumShiftCCRecall{} versus \NumBElevenRecall{}), and under zero-FP calibration the gap widens to \NumShiftCCZeroFPRecall{} versus \NumBElevenZeroFPRecall{}.

\begin{table}[!htbp]
\centering
\scriptsize
\renewcommand{\arraystretch}{0.9}
\caption{Factorial view of the primary setting (FPR / recall at a 1\% budget).}
\label{tab:mechanism}
\begin{tabular}{lcc}
\toprule
Representation & Two-class decision rule & Phishing-only retrieval rule \\
\midrule
TF-IDF & 22.9\% / 98.2\% (B3) & 4.5\% / 79.8\% (B11) \\
Neural embedding & 20.2\% / 99.7\% (B9) & \textbf{5.4\% / 86.4\%} (CyberCane) \\
\bottomrule
\end{tabular}

\end{table}

\begin{figure}[!htbp]
    \centering
    \includegraphics{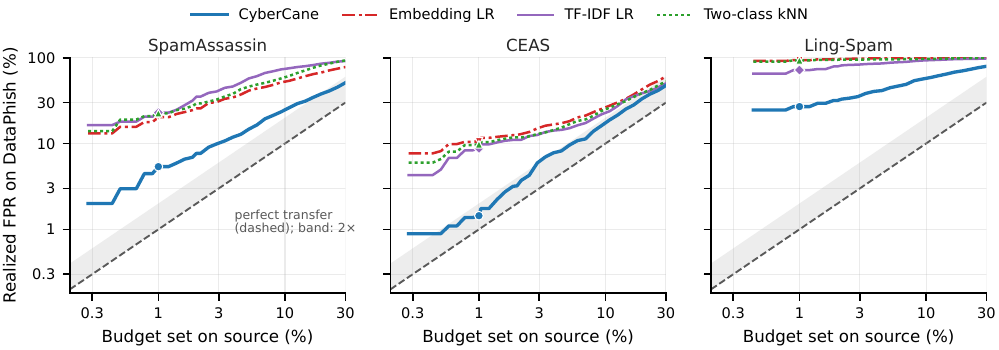}
    \caption{Calibration transfer across budgets. Each curve plots the FPR realized on DataPhish 2025 against the budget its threshold was set for on the source's validation legitimate mail (mean of five seeds; markers: 1\% budget, as in Fig.~\ref{fig:fpr_budget}a). Dashed: perfect transfer; shaded: within 2$\times$ of the budget. Each source is drawn only for budgets its validation set can resolve (budget $\times$ validation size $\geq 1$).}
    \label{fig:transfer}
\end{figure}

\textbf{Transfer across budgets.} Fig.~\ref{fig:transfer} extends the 1\% analysis to every budget a deployer might set on the source, plotting the FPR realized on DataPhish 2025 against it; perfect transfer is the diagonal. No method transfers at any budget: every curve lies above the diagonal throughout. The trained detectors lie far above it, and the gap to CyberCane is widest at the strict budgets that privacy-critical deployments need. At the smallest budget that the \NumToneNvalLo--\NumToneNvalHi{} validation legitimate emails per source can resolve (\NumToneBminLo--\NumToneBminHi), the better trained baseline realizes \NumToneBaseLo--\NumToneBaseHi$\times$ its budget, against \NumToneCCLo--\NumToneCCHi$\times$ for CyberCane; at a 20\% budget its FPR is only \NumToneLenLo--\NumToneLenHi$\times$ CyberCane's, consistent with the crossing frontier of Fig.~\ref{fig:fpr_budget}b.

\textbf{Sensitivity to the score weights.} The neighbor weight and the Phase~1 weight in Eq.~(\ref{eq:risk}) were set before this study and not tuned on its data. Over a grid of \NumWoneConfigs{} settings (neighbor weight 0.4--1.0; Phase~1 weight 0--0.45, where 0 is retrieval alone), CyberCane has lower FPR than both trained baselines in all five seeds of every source and every setting. Even the least favorable setting reduces the overrun \NumWoneFacLo--\NumWoneFacHi$\times$. The defaults rank \NumWoneDefaultRank{}th of \NumWoneConfigs{} on partial AUROC in every source, so the reported results do not rest on a favorable choice of weights.

\textbf{Ranking or calibration?} A budget overrun can arise because a method ranks the deployment mail poorly, or because a threshold set on the source does not transfer. Table~\ref{tab:frontier} separates the two with an oracle analysis that chooses thresholds on the deployment mail itself, which no deployer can do, and reports recall at fixed realized FPR. Under this oracle the trained baselines rank DataPhish well (two-class $k$NN reaches an AUROC of \NumFrBTenAUROC{}; Fig.~\ref{fig:fpr_budget}b), so their \NumRtwoBaseMinOver--\NumRtwoBaseMaxOver$\times$ overrun is a failure of calibration transfer, not of ranking. With the cloud embedder, CyberCane ranks equally well at strict operating points, with the highest partial AUROC at FPR $\leq 2\%$ among the trained detectors and the first zero-shot LLM in all three settings (\NumFrCCpAUCLo--\NumFrCCpAUCHi{}), narrowly ahead of the two-class $k$NN, and its threshold transfers far better (an overrun \NumRtwoFacLo--\NumRtwoFacHi$\times$ smaller); a larger LLM ranks higher (Section~\ref{sec:llm}). With a local embedder the two-class $k$NN ranks better (Table~\ref{tab:c6_local}), so this ranking advantage depends on the embedding model, whereas the calibration advantage does not. The two-class methods, in turn, rank better at lenient operating points, consistent with their training on both classes.

\begin{table}[!htbp]
\centering
\scriptsize
\renewcommand{\arraystretch}{0.9}
\caption{Ranking quality on the deployment mail (oracle thresholds chosen on DataPhish 2025 labels; descriptive, not deployable). Recall at fixed realized FPR, calibration source SpamAssassin, mean of five seeds; pAUC is McClish-normalized.}
\label{tab:frontier}
\begin{tabular}{lcccccccc}
\toprule
 & \multicolumn{5}{c}{Recall (\%) at realized FPR} & \multicolumn{2}{c}{pAUC$_{\leq 2\%}$} & \\
\cmidrule(lr){2-6}\cmidrule(lr){7-8}
Method & 0.5\% & 1\% & 2\% & 5\% & 10\% & SpamAssassin & all sources & AUROC \\
\midrule
\textbf{CyberCane} & 64.4 & 73.2 & 82.3 & 86.4 & 90.4 & 0.863 & 0.850--0.864 & 0.954 \\
Zero-shot LLM (B12) & 46.4 & 56.9 & 65.7 & 89.9 & 96.6 & 0.787 & 0.754--0.787 & 0.979 \\
Two-class $k$NN (B10) & 59.9 & 73.5 & 84.7 & 91.3 & 99.3 & 0.859 & 0.842--0.859 & 0.989 \\
Embedding LR (B9) & 48.9 & 65.7 & 75.4 & 86.7 & 98.8 & 0.808 & 0.808--0.819 & 0.984 \\
TF-IDF LR (B3) & 63.5 & 67.8 & 72.8 & 85.2 & 92.6 & 0.825 & 0.759--0.838 & 0.977 \\
Phishing-only TF-IDF $k$NN (B11) & 51.1 & 65.0 & 73.4 & 84.0 & 92.1 & 0.804 & 0.785--0.806 & 0.960 \\
\bottomrule
\end{tabular}

\end{table}

Table~\ref{tab:style_shift} separates the two candidate causes of shift. When legitimate mail comes from the same distribution and only the attack author changes, from human-written to LLM-written phishing of disjoint families, no method exceeds \NumRfiveMaxFPR{} FPR at the 1\% budget, so the budget overrun of Section~\ref{sec:shift} stems from benign-distribution shift. For every embedding-based and trained method, recall on LLM-written phishing matches the human-to-human control within about two points. The exception shows why embeddings matter: the phishing-only TF-IDF rule loses recall on LLM rewrites and recovers only \NumRfiveBElevenRecall{} of them, against \NumRfiveCCRecall{} for CyberCane, consistent with rewrites that change the wording but not the content. With a stable benign distribution, CyberCane has no FPR advantage and trails the trained baselines by four to six points of recall.

\begin{table}[!htbp]
\centering
\scriptsize
\renewcommand{\arraystretch}{0.9}
\caption{Attack-style shift within DataPhish 2025 (1\% budget; mean$\pm$SD over five seeds). Legitimate mail comes from the same distribution in training and test.}
\label{tab:style_shift}
\begin{tabular}{lcccc}
\toprule
 & \multicolumn{2}{c}{Human $\rightarrow$ LLM-written} & \multicolumn{2}{c}{Human $\rightarrow$ human (control)} \\
\cmidrule(lr){2-3}\cmidrule(lr){4-5}
Method & FPR (\%) & Recall & FPR (\%) & Recall \\
\midrule
TF-IDF LR (B3) & 1.2$\pm$0.8 & 91.7\% & 0.5$\pm$0.4 & 89.8\% \\
Embedding LR (B9) & 0.9$\pm$0.6 & 92.6\% & 0.6$\pm$0.4 & 91.6\% \\
Two-class $k$NN (B10) & 0.5$\pm$0.2 & 93.4\% & 0.4$\pm$0.2 & 92.5\% \\
Phishing-only TF-IDF $k$NN (B11) & 0.5$\pm$0.4 & 27.2\% & 0.5$\pm$0.5 & 44.0\% \\
\textbf{CyberCane} & 0.9$\pm$0.5 & 87.7\% & 1.1$\pm$0.6 & 87.6\% \\
\bottomrule
\end{tabular}

\end{table}

\subsection{Segment-Level Scoring: Padding Robustness and Threshold Transfer}
\label{sec:segment}

Because CyberCane scores only resemblance to known attacks, an attacker might append benign text so that the whole-email embedding resembles legitimate mail. We tested this on DataPhish 2025 by appending legitimate DataPhish training text to every test phishing email at 0.5, 1, and 2 times its length, keeping all detectors and thresholds as calibrated on clean source data. Padding of equal length cuts the recall of whole-email CyberCane to \NumFullPadLo--\NumFullPadHi{} (Table~\ref{tab:segment}).

Segment-level scoring removes this weakness. CyberCane-Seg keeps \NumSegRetLo--\NumSegRetHi{} of its recall under equal-length padding, against \NumFullRetLo--\NumFullRetHi{} for whole-email scoring and more than the best trained detector in two of three settings. Its recall barely changes as the padding doubles, because the phishing segments remain phishing-like however much benign text surrounds them. On clean mail it ranks the deployment mail as well as whole-email CyberCane (partial AUROC within \NumSegPAUCDiff{}), yet its source-calibrated threshold transfers better: the realized FPR at the 1\% budget falls from \NumSegFPRSAFrom{} to \NumSegFPRSATo{} (SpamAssassin), \NumSegFPRCEASFrom{} to \NumSegFPRCEASTo{} (CEAS), and \NumSegFPRLingFrom{} to \NumSegFPRLingTo{} (Ling-Spam). Operating closer to the budget costs \NumSegRecCostLo--\NumSegRecCostHi{} points of recall at that threshold.

\begin{table}[!htbp]
\centering
\scriptsize
\renewcommand{\arraystretch}{0.9}
\caption{Whole-email versus segment-level scoring on DataPhish 2025 (mean of five seeds). Clean: realized FPR and recall at the 1\% budget. Padded: recall when every phishing email is padded with an equal length of legitimate text, at the clean threshold. Oracle: partial AUROC on clean mail.}
\label{tab:segment}
\begin{tabular}{llcccc}
\toprule
 & & \multicolumn{2}{c}{Clean, 1\% budget} & Padded ($r{=}1$) & Oracle \\
\cmidrule(lr){3-4}\cmidrule(lr){5-5}\cmidrule(lr){6-6}
Source & Scoring & FPR \% & Recall \% & Recall \% & pAUC$_{\leq 2\%}$ \\
\midrule
SpamAssassin & CyberCane & 5.4 & 86.4 & 51.7 & 0.863 \\
 & CyberCane-Seg & 2.7 & 82.0 & 71.7 & 0.862 \\
\addlinespace
CEAS & CyberCane & 1.4 & 78.4 & 24.9 & 0.850 \\
 & CyberCane-Seg & 1.2 & 77.1 & 64.1 & 0.852 \\
\addlinespace
Ling-Spam & CyberCane & 26.9 & 94.7 & 85.4 & 0.864 \\
 & CyberCane-Seg & 9.9 & 88.9 & 83.6 & 0.863 \\
\bottomrule
\end{tabular}

\end{table}

\subsection{Can Unlabeled Deployment Mail Repair Calibration?}
\label{sec:recalibration}

If trained detectors fail by calibration transfer, an organization might recalibrate them on its own unlabeled incoming mail. We test the most direct label-free remedy: fit a two-component Gaussian mixture to the scores of unlabeled deployment mail, take the lower component as legitimate, and place the threshold at its $(1-b)$ quantile. The assumption is that the scores of unlabeled mail roughly form two groups, so the lower group can stand in for the legitimate mail that has no labels. As a generous reference we also give each method the true phishing prevalence of the unlabeled pool, which no deployer knows.

On DataPhish (Table~\ref{tab:r6_recalibration}), mixture recalibration narrows the trained baselines' overrun but leaves them at \NumRsixBaseLo--\NumRsixBaseHi{} FPR under SpamAssassin and CEAS calibration, whereas CyberCane recalibrated the same way meets the 1\% budget for every calibration source (at most \NumRsixCCFPR{} FPR), at a recall of \NumRsixCCRecall{}. Knowing the true prevalence does not help: it leaves every method, CyberCane included, at \NumRsixLoneLo--\NumRsixLoneHi{} FPR. The result does not generalize to the older-corpus shifts of Table~\ref{tab:b2_generality}, where the same recalibration largely repairs the trained baselines (best baseline \NumBtwoLtwoBaseLo--\NumBtwoLtwoBaseHi{}) and CyberCane is not better (\NumBtwoLtwoCCLo--\NumBtwoLtwoCCHi{}). Label-free recalibration is therefore a remedy worth applying to any detector; CyberCane's advantage under it appeared on DataPhish but not on the older corpora, so it should not be assumed in general.

\begin{table}[!htbp]
\centering
\scriptsize
\renewcommand{\arraystretch}{0.9}
\caption{Label-free recalibration on unlabeled DataPhish mail (FPR / recall in \%, 1\% budget, mean of five seeds). Oracle prevalence: thresholds set with the true phishing share of the unlabeled pool, unavailable in practice.}
\label{tab:r6_recalibration}
\begin{tabular}{lcccc}
\toprule
 & \multicolumn{2}{c}{CyberCane} & Best trained baseline & CyberCane \\
\cmidrule(lr){2-3}
Calibration source & Source-calibrated & Label-free mixture & Label-free mixture & Oracle prevalence \\
\midrule
SpamAssassin & 5.4 / 86.4 & \textbf{0.3 / 62.1} & 5.0 / 91.7 (Two-class $k$NN) & 7.2 / 88.2 \\
CEAS & 1.4 / 78.4 & \textbf{0.3 / 62.0} & 8.3 / 96.4 (Two-class $k$NN) & 7.6 / 88.2 \\
Ling-Spam & 26.9 / 94.7 & \textbf{0.3 / 62.1} & 0.8 / 66.0 (Two-class $k$NN) & 7.0 / 88.3 \\
\bottomrule
\end{tabular}

\end{table}

\subsection{Comparison With Zero-Shot LLMs}
\label{sec:llm}

A zero-shot LLM learns nothing from the calibration corpus and, like CyberCane, keeps close to its budget under benign shift (Table~\ref{tab:main_shift}: \NumLLMFPR{} FPR for gpt-oss-120b). How it ranks depends strongly on the model (Table~\ref{tab:llm_compare}). Against gpt-oss-120b the oracle frontiers cross: CyberCane recovers more phishing at strict operating points (\NumFrCCOne{} versus \NumFrLLMOne{} at 1\% realized FPR), the LLM at lenient ones (\NumFrLLMTen{} versus \NumFrCCTen{} at 10\%). Llama 3.3 70B ranks better than CyberCane throughout, recovering \NumLlamaOneLo{} at 1\% realized FPR against \NumCCOneLo--\NumCCOneHi{}, and stays within budget at its fixed cut-off (\NumLlamaDefaultFPR{} FPR, \NumLlamaDefaultRecall{} recall). The two designs therefore occupy different points of the deployment trade-off. CyberCane's verdict needs one embedding and an indexed lookup: measured on the same emails, it takes \NumLatLocal{}\,s with a local embedder and \NumLatCloud{}\,s with the cloud embedder, against \NumLatLLM{}\,s for the 70B LLM, costs nothing or \$\NumCostCloud{} per million emails against \$\NumCostLLM{}, and sends either no email text or a redacted embedding input to a third party, whereas the LLM must receive the redacted email in its prompt. Its verdict is also traceable to retrieved attacks and rules.

\begin{table}[!htbp]
\centering
\scriptsize
\renewcommand{\arraystretch}{0.9}
\caption{Ranking quality and measured deployment cost. Ranking: oracle recall at 1\% realized FPR and partial AUROC on DataPhish 2025, range over the three calibration sources (five seeds). Per email: median latency over 100 DataPhish emails (8 CPU threads; LLM: 50 sequential calls), cost per million emails (embedding list price; LLM: mean recorded cost of all study calls), and characters of redacted email text sent to an external service. CyberCane-Seg adds a few short embeddings to the cloud configuration.}
\label{tab:llm_compare}
\begin{tabular}{lccccc}
\toprule
 & \multicolumn{2}{c}{Ranking (oracle, 3 sources)} & \multicolumn{3}{c}{Per email (measured)} \\
\cmidrule(lr){2-3}\cmidrule(lr){4-6}
Method & Recall \% at 1\% FPR & pAUC$_{\leq 2\%}$ & Latency (s) & USD per $10^6$ & Chars sent out \\
\midrule
CyberCane, local embedder & \multicolumn{2}{c}{lower; Table~\ref{tab:c6_local}} & 0.20 & 0.00 & 0 \\
CyberCane, cloud embedder & 71.4--73.2 & 0.850--0.864 & 0.33 & 6.45 & 788 \\
CyberCane-Seg, cloud embedder & 72.1--73.8 & 0.852--0.863 & -- & -- & -- \\
Zero-shot gpt-oss-120b (B12) & 51.6--56.9 & 0.754--0.787 & -- & 33.26 & 1362 \\
Zero-shot Llama 3.3 70B (B13) & 83.7 & 0.918--0.926 & 4.62 & 63.67 & 1362 \\
\bottomrule
\end{tabular}

\end{table}

 The gpt-oss LLM is, however, more robust to an unrepresentative calibration source (\NumLLMLingFPR{} FPR with Ling-Spam, where CyberCane is not; Table~\ref{tab:llm_frontier}). Combining both scores (an average of their calibrated ranks) improved recall and FPR over CyberCane on the development source. It met the preregistered criterion on neither confirmation source (Table~\ref{tab:llm_frontier}), although it halved the FPR under the narrow Ling-Spam calibration, so we report it as a supplementary result. With Llama 3.3 70B, neither this combination nor the cascade below improved on the LLM alone in a majority of settings. A second combination sends only mail above CyberCane's review threshold to the LLM and raises an alert only when both agree. It lowered CyberCane's FPR under Ling-Spam calibration from \NumVoneLingFrom{} to \NumVoneLingTo{} and had lower FPR than the same cascade built on any trained baseline, but realized \NumVoneFPRLo--\NumVoneFPRHi{} FPR overall and so missed its preregistered 2\% bound. The remaining false positives are legitimate messages that both stages flag (the LLM alone flags \NumVoneLLMLo--\NumVoneLLMHi{} of DataPhish legitimate mail at its fixed cut-off), and \NumVoneEscLo--\NumVoneEscHi{} of legitimate mail would reach the external model.

\begin{table}[!htbp]
\centering
\scriptsize
\renewcommand{\arraystretch}{0.9}
\caption{CyberCane, the zero-shot LLM (B12), and their score fusion across calibration sources (1\% budget, mean of five seeds).}
\label{tab:llm_frontier}
\begin{tabular}{lccc}
\toprule
 & \multicolumn{3}{c}{FPR (\%) / recall (\%) at a 1\% budget} \\
\cmidrule(lr){2-4}
Calibration benign source & CyberCane & Zero-shot LLM & Fusion \\
\midrule
SpamAssassin (development) & 5.4 / 86.4 & 0.7 / 56.4 & 3.8 / 92.4 \\
CEAS (confirmation) & 1.4 / 78.4 & 0.8 / 43.9 & 1.6 / 80.4 \\
Ling-Spam (confirmation) & 26.9 / 94.7 & 1.3 / 65.7 & 12.6 / 95.9 \\
\bottomrule
\end{tabular}

\end{table}

\subsection{System Evidence: Privacy Boundary, Ontology, and Explanations}
\label{sec:system}

\textbf{Privacy boundary.} Regression tests intercept every outbound request of a full analysis and confirm that planted names, addresses, SSNs, phone numbers, dates of birth, record numbers, and email addresses do not reach the embedding or language-model services. Across the \NumPrivRows{} emails sanitized for this study, Table~\ref{tab:privacy} counts residual identifier patterns after redaction. Apart from publication dates inside news-article URL paths (\NumPrivResidualDates{} rows), one documentation placeholder address, and one phone number inside a phone-lookup URL, the residue is IPv4 addresses. The redactor used in all experiments (version 2.2) has no IP rule, and \NumPrivResidualIP{} emails (\NumPrivResidualIPPct{}) keep one in their text. Pattern counts cannot measure recall, and anonymization should be judged both by the identifiers it misses and by the content it masks needlessly~\citep{pilan2022tab}. We therefore planted synthetic identifiers with known ground truth into 200 test emails, once after a cue (``My SSN is \dots'') and once without, and scored whether any part of each survived (Table~\ref{tab:privacy_planted}; \NumPlantN{} plantings, \NumPlantLost{} excluded because the named-entity model also tagged the markers delimiting the planted text). Six structured types were removed in every case. Version 2.2 missed day-first dates (\NumPlantDobDMY{} recall, against \NumPlantDobOther{} for three other formats), house numbers on streets with two unlisted suffixes, and every IP address. Version 2.3 adds rules for all three and removes them on the same audit; this estimate is not independent, because the rules were written after the audit. Person names are the weak point: recall is \NumPlantNameCued{} after a cue and \NumPlantNameUncued{} without one, ranging without a cue from \NumPlantNameLo{} for African to \NumPlantNameHi{} for Arabic first names (20--45 names per group), a disparity we attribute to the named-entity model. Brand names, URL hosts, and reply domains were kept in every case. Identifiers planted in fixed templates are easier to find than natural ones, so these rates are likely optimistic for natural text. The preregistered human audit of natural identifiers (TB; Appendix~\ref{app:annotation}; 28 emails after two were flagged as not judgeable, 91 checkable identifiers) does not meet its criterion for calling personal identifiers largely removed: 8 of 25 identifiers classified as personal leak under the planted-audit rule (32\%, CI 12.5--56.3). On inspection, seven of the eight are common words (such as ``details'' and ``below'') or mailing-list names that the redactor masked and the annotator classified as personal, and that the rule counts as leaked because the word recurs elsewhere in the email; the eighth is the year of a date that also appears in another format. Of the 155 placeholders, 23\% hid a harmless word, and none of these was judged to make the email harder to assess. Version~2.3 leaks the same identifiers.

\begin{table}[!htbp]
\centering
\scriptsize
\renewcommand{\arraystretch}{0.9}
\caption{Residual identifier patterns in sanitized text (all corpora used in this study).}
\label{tab:privacy}
\begin{tabular}{lr}
\toprule
Residual pattern after redaction & Rows \\
\midrule
Email address (local part) & 1 \\
US SSN & 0 \\
Phone number & 1 \\
Digit run $\geq$6 (IDs, accounts) & 0 \\
Numeric date & 22 \\
IPv4 address (outside URLs) & 690 \\
\midrule
Rows audited & 26,075 \\
\bottomrule
\end{tabular}

\end{table}

\begin{table}[!htbp]
\centering
\scriptsize
\renewcommand{\arraystretch}{0.9}
\caption{Planted-identifier audit: recall (\%) of the privacy boundary on synthetic identifiers planted in 200 test emails, after a contextual cue or without one (about 195 plantings per cell; at 100\% recall the lower 95\% Wilson bound is 98\%). All experiments used version 2.2; version 2.3 adds rules for the failures found here, so its columns are not an independent estimate.}
\label{tab:privacy_planted}
\begin{tabular}{lcccc}
\toprule
 & \multicolumn{2}{c}{Redactor v2.2} & \multicolumn{2}{c}{Redactor v2.3} \\
\cmidrule(lr){2-3}\cmidrule(lr){4-5}
Identifier type & Cued & Uncued & Cued & Uncued \\
\midrule
Email address & 100.0 & 100.0 & 100.0 & 100.0 \\
Phone number & 100.0 & 100.0 & 100.0 & 100.0 \\
US SSN & 100.0 & 100.0 & 100.0 & 100.0 \\
Payment card & 100.0 & 100.0 & 100.0 & 100.0 \\
Account number & 100.0 & 100.0 & 100.0 & 100.0 \\
Medical record no. & 100.0 & 100.0 & 100.0 & 100.0 \\
Date of birth & 73.8 & 77.3 & 100.0 & 100.0 \\
IPv4 address & 0.0 & 0.0 & 100.0 & 100.0 \\
Person name & 94.4 & 79.8 & 93.9 & 79.8 \\
Street address & 77.2 & 83.0 & 100.0 & 100.0 \\
\midrule
\multicolumn{5}{l}{Phishing signals kept (v2.2): brand 100\%, URL host 100\%, reply domain 100\%} \\
\bottomrule
\end{tabular}

\end{table}

\textbf{Authentication under live DNS.} The public corpora carry no headers, so the authentication rules are not exercised in the main experiments. We therefore scored \NumAuthN{} legitimate and \NumAuthPhishN{} phishing emails from the test splits with live DNS lookups of their sender domains. A literal reading of the records, which requires an MX record and a DMARC record on the exact sender domain, flagged \NumAuthStrict{} legitimate emails as phishing, mostly mail from subdomains of organizations with valid records. The standards-conformant checks of Section~\ref{sec:phase1} flagged \NumAuthRfc{} (\NumAuthRedPct{} fewer) while still flagging all \NumAuthPhishKept{} phishing emails. Of the remaining alarms, \NumAuthDead{} come from sender domains that no longer exist, a red flag for live mail that here reflects the age of the corpora.

\textbf{Ontology.} All \NumOntRules{} attack classes are parsed from the ontology's axioms (Table~\ref{tab:ontology_rules}). A single indicator no longer yields attack labels; it yields suggestions with explicit missing premises. For example, a credential request alone suggests \texttt{CredentialTheft} with the missing premise \texttt{hasMissingMX}, rather than asserting it.

\begin{figure}[!htbp]
    \centering
    \includegraphics[width=\textwidth]{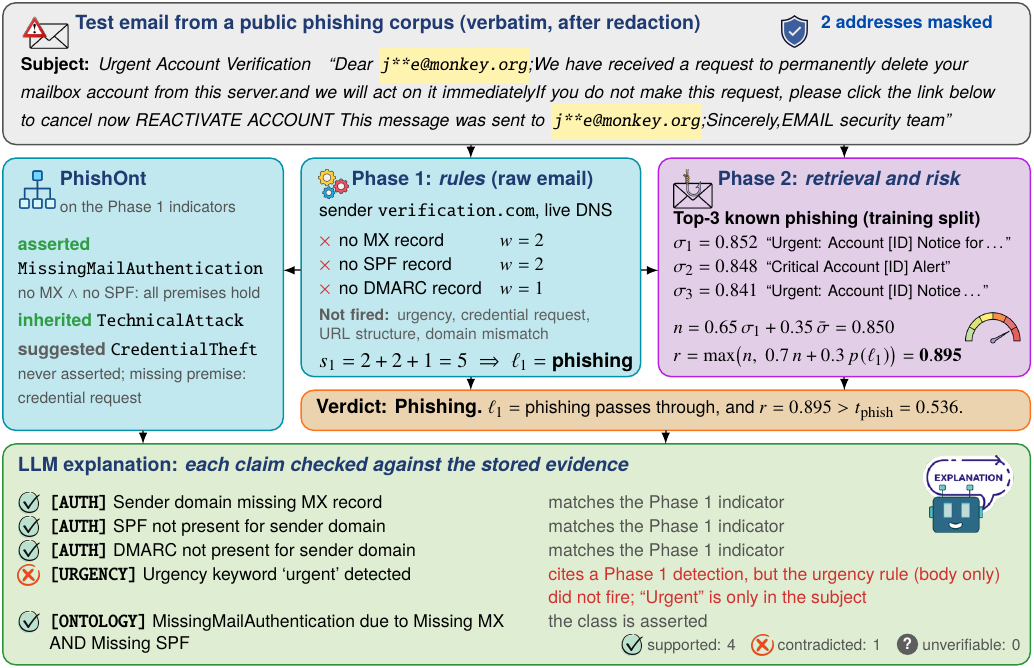}
\caption{A checked explanation for a phishing test email from the explanation audit (real system output; claims shortened). Phase~1 analyzes the raw email; retrieval and the explanation model see only the redacted text. Each claim is checked against the stored evidence, and the verdict is fixed before the explanation is generated.}
    \label{fig:explanation_example}
\end{figure}

\textbf{Explanations.} Table~\ref{tab:groundedness} evaluates the explanation checker on a challenge set of \NumChallengeWrong{} plausible but wrong explanations, including invented similarity values, similarity claimed with no retrieved neighbors, unasserted attack classes, and quotes absent from the email. A keyword-based checker of the kind used in an earlier version of this work accepts all \NumChallengeVoneSim{} invented similarity claims; this explains the uniformly perfect similarity-support rates such a checker reports. The evidence-based checker accepts \NumChallengeVtwo{} of the wrong explanations and supports \NumChallengeRight{} faithful ones. In the preregistered human audit (E1; Appendix~\ref{app:annotation}; 60 emails, 260 sentences of the deployed variant), the annotator labeled 69.2\% of sentences supported (95\% CI 61.5--76.5) and 28.8\% contradicted (21.6--36.1), so neither the ``largely grounded'' nor the ``rarely wrong'' criterion is met. Ontology context made no detectable difference (support $+4.2$ points, CI $-2.7$ to $10.7$). A post hoc check shows that these rates largely measure annotation error. Of the 260 sentences, 213 restate a field of the evidence panel (a sender record, a similarity value, an empty list of warning signs) and can be checked mechanically; 212 are true, yet the annotator labeled 60 of them contradicted, mostly true statements that a record is present, a sentence type the quality gate did not cover. Using the mechanical truth for these sentences, support is 92.7\% (88.9--96.0) and contradiction 6.2\% (3.1--9.7); on the 47 sentences that need judgment, support is 61.7\% (46.0--77.1), with most contradictions in urgency and ontology claims. The automated checker's precision against the human labels is 0.94 (preregistered threshold 0.90) at 69\% coverage, and the annotator rated 96 of 119 explanations helpful and 113 consistent with the verdict. Fig.~\ref{fig:explanation_example} traces one email from this audit. Under live DNS, three authentication rules fire ($s_1=5$), so the Phase~1 verdict passes through, and the risk score agrees ($r=0.895$). PhishOnt asserts \texttt{MissingMailAuthentication} and its superclass but only suggests \texttt{CredentialTheft}, because no credential request was found. Of the five generated claims, the checker supports four and contradicts one, which credits the keyword ``urgent'' to Phase~1. The urgency rule reads only the body, and the word occurs only in the subject. Unlike the challenge set, this error was not planted. The case also exposes a gap in the rules: urgency in the subject line is not scored.

\begin{table}[!htbp]
\centering
\scriptsize
\renewcommand{\arraystretch}{0.9}
\caption{Explanation-checker challenge set: wrong explanations judged supported by a keyword checker (v1) and by the evidence-based checker (v2).}
\label{tab:groundedness}
\begin{tabular}{lcccc}
\toprule
 & Wrong & \multicolumn{2}{c}{Wrong judged supported} & Faithful judged \\
\cmidrule(lr){3-4}
Tag & cases & v1 (keyword) & v2 (evidence) & supported (v2) \\
\midrule
SIMILARITY & 5 & 5 & 0 & 2 / 2 \\
ONTOLOGY & 5 & 0 & 0 & 1 / 1 \\
AUTH & 6 & 0 & 0 & 2 / 2 \\
URL & 7 & 0 & 0 & 1 / 1 \\
URGENCY & 5 & 0 & 0 & 2 / 2 \\
CONTENT & 2 & 0 & 0 & 1 / 1 \\
\midrule
Total & 30 & 5 & 0 & 9 / 9 \\
\bottomrule
\end{tabular}

\end{table}

\subsection{Negative Results}
\label{sec:negative}

Besides the score fusion of Section~\ref{sec:llm}, two preregistered attempts to extend the approach did not meet their criteria (Table~\ref{tab:negative}). A label-free CSLS-style correction, which subtracts each email's mean similarity to unlabeled mail of its own domain~\citep{conneau2018word}, reduced the FPR only for the narrow Ling-Spam calibration and lowered low-FPR ranking quality everywhere, plausibly because unlabeled pools also contain phishing. Few-shot adaptation inserted up to 50 labeled new-era phishing emails into the index without retraining. It raised recall by at most \NumRfourMaxGain{} points, consistent with the missed test emails resembling none of the added examples, since candidates near-duplicating test emails were excluded. Retraining the baselines with the same examples did not remove their overrun either. Finally, on the phishing-or-spam task CyberCane stays within budget (\NumUnwantedCCFPR{} FPR) but at low recall (\NumUnwantedCCRecall{}), because spam is a poor fit for a phishing index. We report these results because they bound the claims above.

\begin{table}[!htbp]
\centering
\scriptsize
\renewcommand{\arraystretch}{0.9}
\caption{Negative results: label-free CSLS correction (E2) and few-shot index adaptation (R4), at a 1\% budget.}
\label{tab:negative}
\begin{tabular}{lcccc}
\toprule
 & \multicolumn{3}{c}{Label-free CSLS correction (E2)} & Few-shot index adaptation (R4) \\
\cmidrule(lr){2-4}\cmidrule(lr){5-5}
Benign source & Retrieval FPR/recall (\%) & CSLS FPR/recall (\%) & pAUC$_{\leq 2\%}$ & Recall, $k{=}0 \rightarrow 50$ (\%) \\
\midrule
SpamAssassin & 6.0 / 87.8 & 6.0 / 89.2 & 0.862 $\rightarrow$ 0.667 & 86.4 $\rightarrow$ 87.1 \\
CEAS & 2.2 / 82.2 & 3.3 / 53.3 & 0.850 $\rightarrow$ 0.636 & 78.4 $\rightarrow$ 78.9 \\
Ling-Spam & 36.1 / 95.8 & 10.5 / 93.1 & 0.863 $\rightarrow$ 0.664 & 94.7 $\rightarrow$ 94.7 \\
\bottomrule
\end{tabular}

\end{table}

\FloatBarrier
\section{Discussion}
\label{sec:discussion}

\textbf{Implications for deployment.} The main practical lesson is that an FPR budget set on public data should be checked on an organization's own mail before it is relied on: realized FPR can exceed the budget by one to two orders of magnitude, even with conservative calibration. Because the issue lies in calibration transfer rather than ranking, held-out accuracy on the training corpus does not reveal it, whereas a measurement on deployment mail does. A detector whose positive decisions depend on resemblance to known attacks, rather than on a learned boundary around another corpus's legitimate mail, keeps the overrun much smaller, provided it uses representations that generalize to new attack wording. Where unlabeled deployment mail is available, a label-free mixture recalibration is worth applying to any detector (Section~\ref{sec:recalibration}). On DataPhish it brought CyberCane within budget where the trained baselines remained over it; on the older corpora it repaired the trained baselines as well. Once labeled legitimate mail becomes available, recalibrating on it is worthwhile, since a representative calibration set benefits every method, CyberCane included.

\textbf{Choosing among designs.} When labeled legitimate mail from the deployment site exists, trained embedding classifiers are the strongest choice: they stay near budget and reach the highest recall (Table~\ref{tab:in_domain}). When it does not, the choice depends on what may leave the organization and what a verdict may cost. A large zero-shot LLM ranks best and is insensitive to the calibration source, and it requires each email to be sent to it, at about \NumSpeedLo--\NumSpeedHi{} times CyberCane's latency. CyberCane keeps the decision local, at \NumLatLocal{}\,s and no per-email cost, with its segment-level score staying closest to the budget among the calibrated detectors, and every verdict traceable to specific retrieved attacks, rules, and asserted ontology classes. Where a cloud LLM is acceptable, CyberCane can still screen all mail locally and reserve the LLM for cases that need it.

\textbf{Privacy.} The boundary minimizes and measures what leaves the organization. A cloud embedding service still receives redacted text, and redacted embeddings can be partly inverted~\citep{li2023sentence}; an open embedding model run locally removes external transmission for detection altogether. In that configuration the calibration advantage holds, with \NumCsixFacLo--\NumCsixFacHi$\times$ lower FPR than the better trained baseline, with \NumCsixRecLossLo--\NumCsixRecLossHi{} fewer recall points at the 1\% budget (Table~\ref{tab:c6_local}), so organizations can choose between maximum privacy and maximum recall.

\textbf{Limitations.} Each conclusion rests on an assumption, and naming them shows where further evidence would help most. We assume that older public collections stand in for an organization's legitimate mail; where real traffic differs more from the calibration source, as Ling-Spam already does, realized FPR rises for every method, CyberCane included, which is why calibrating on representative mail pays off (Table~\ref{tab:benign_sources}). We assume that DataPhish rewrites, which are paraphrases of human originals, represent new attacks; beyond benign padding, which segment-level scoring handles, adaptive attacks on the retrieval index remain to be tested, and their effect on recall is not yet measured. The public corpora carry no full headers, so Phase~1's authentication checks were exercised only on a live-DNS sample, against today's records rather than those at sending time. Redaction recall was measured on planted identifiers, which are easier to find than natural ones, and the named-entity model recognizes uncued names of some origins much better than others, which makes it the component most worth improving. Results rely on two embedding models and two zero-shot LLMs; CyberCane's strict-end ranking advantage holds against the smaller LLM and with one of the two embedders, so its main strengths are calibration, locality, and cost. Explanations and natural-text redaction were audited by a single annotator who is one of the authors (Appendix~\ref{app:annotation}), whose labels disagree with the evidence on 122 of 403 mechanically checkable sentences (post hoc); a second annotator would most strengthen these audits.

\FloatBarrier
\section{Conclusion}
\label{sec:conclusion}

Calibrating a phishing detector on public corpora and deploying it elsewhere, the natural choice when privacy prevents collecting labeled legitimate mail, leads trained detectors to exceed their false-positive budgets by one to two orders of magnitude. The cause is calibration transfer, not ranking, and this points to a direct remedy. CyberCane, which scores only by similarity to known phishing in a redacted, training-only index, removes legitimate mail from the score so that only its threshold depends on the calibration source. It reduces the overrun \NumRtwoFacLo--\NumRtwoFacHi$\times$ across three calibration sources and five seeds, on a second family of shifts that does not involve DataPhish, and with an open, locally run embedder; scoring the most phishing-like segment lowers the realized FPR a further \NumSegOverRedLo--\NumSegOverRedHi$\times$ and resists benign padding. It screens an email locally in \NumLatLocal{}\,s, where a stronger 70B LLM needs \NumLatLLM{}\,s and the email itself. Neural embeddings are what keep its recall on unseen attacks. The shift in legitimate mail, not LLM authorship, drives the overrun; semantic retrieval keeps its recall under LLM rewriting, which affects only lexical retrieval; and a group-aware split removes the leakage through which a random split of a current benchmark exposes most of its test set. Promising next steps are narrowing the recall gap of local embedders, recalibrating from small amounts of site-specific legitimate mail, and extending the padding defense to adaptive attacks on the retrieval index.

\FloatBarrier
\section*{Statements and Declarations}

\textbf{Funding.} \CCFunding

\textbf{Conflict of interest.} The authors declare that there is no conflict of interest.

\textbf{Data availability.} All corpora are public~\citep{nazario2007phishing,spamassassin2006corpus,toth2025dataphish,androutsopoulos2000lingspam,alsubaiey2024phishing}. Split manifests (row hashes, group identifiers, and counts), per-example predictions, and run manifests for every experiment are released with the code.

\textbf{Code availability.} The CyberCane implementation, the privacy boundary, PhishOnt, the evaluation harness, and the scripts that generate every table and figure are available at \url{https://github.com/sbhakim/Cybercane}.

\textbf{Ethics approval.} This study used only publicly available email corpora. The human annotation (Appendix~\ref{app:annotation}) was performed by one of the authors and collected labels on system outputs, not data about the annotator. The study did not involve human participants or animals, and no personal data were collected; ethics approval was therefore not required.

\textbf{Author contributions.} Conceptualization, S.B.H., A.A., Q.Z., V.M. and P.S.; methodology, S.B.H. and A.A.; software, S.B.H. and P.S.; validation, S.B.H. and A.A.; formal analysis, S.B.H.; investigation, S.B.H.; data curation, S.B.H.; writing---original draft preparation, S.B.H. and A.A.; writing---review and editing, S.B.H. and A.A.; visualization, S.B.H.; supervision, H.H.S.; project administration, S.B.H. All authors have read and agreed to the published version of the manuscript.

\clearpage\appendix\section{Phase 1 Rules and Redaction Specifications}\label{app:rule-details}

\textbf{Weighted rules.} Missing MX record (2); missing SPF record (2); missing DMARC record (1); lookalike sender or link domain (2); freemail sender claiming a brand (2); IP-literal link (2); shortened link (bit.ly, tinyurl.com, t.co, goo.gl, ow.ly, buff.ly) (2); urgency language (2); credential request (2). DNS-based rules are disabled when DNS is unavailable, as for the historical domains of the public corpora.

\textbf{Content cues.} Urgency: \textit{urgent}, \textit{immediate action}, \textit{verify your account}, \textit{password expires}, \textit{suspend}, \textit{pay now}, \textit{update your information}. Credential request: mentions of passwords, log-in, SSN, social security, credit cards, or bank accounts.

\textbf{Redaction.} Each identifier is replaced by a typed placeholder: \texttt{[EMAIL]@domain} (the domain is kept as a phishing signal), \texttt{[PHONE]}, \texttt{[SSN]}, \texttt{[CARD]}, \texttt{[DATE]}, \texttt{[ID]} (labeled record, member, account, policy, and claim numbers, and digit runs of six or more), \texttt{[ZIP]}, \texttt{[ADDRESS]}, \texttt{[NAME]}, and \texttt{[LOCATION]} (from the named-entity recognizer, confidence $\geq 0.70$). In URLs, hosts and paths are kept, query strings are replaced by \texttt{?[QUERY]}, userinfo by \texttt{[USERINFO]@}, and digit runs in paths by \texttt{[ID]}. Email matching uses ASCII boundaries so that addresses adjacent to CJK text are detected.

\FloatBarrier
\section{Prompts}\label{app:prompts}

\textbf{Zero-shot baseline (B12).} \textit{``You are an email security analyst. Decide whether the email below is a phishing attempt (it tries to steal credentials, money, or personal information, or to make the recipient click a malicious link or open a malicious attachment) or a legitimate email. Personal identifiers have been replaced with placeholders such as [NAME], [EMAIL], [PHONE], and [ID]; do not treat placeholders themselves as suspicious. Respond with JSON only, in exactly this form: \{"phishing\_probability": <integer 0-100>, "verdict": "phishing" or "legitimate", "reason": "<one sentence>"\} EMAIL:''} followed by the redacted email (at most 3{,}000 characters). Temperature 0; low reasoning effort; one pinned bf16 provider.

\textbf{Explanations.} The explanation prompt supplies the redacted subject and body, the Phase~1 evidence, neighbor statistics (identifier, similarity, redacted subject), and asserted ontology classes, and asks for three to five tagged explanations that cite the supplied evidence. Explanations do not influence the verdict.

\FloatBarrier
\section{Preregistered Experiments}\label{app:prereg}

Table~\ref{tab:prereg} lists every experiment of this study, its preregistered criterion, and its outcome. Protocols and criteria were recorded in a dated log before each run; full records, including run manifests and per-example predictions, are part of the released materials. The primary hypothesis itself was formulated after an exploratory probe in which a TF-IDF classifier overran its budget under shift; that probe is not among the reported results, and every CyberCane result was obtained after its protocol had been fixed. Table~\ref{tab:in_domain} gives the in-domain reference results (training and testing within one corpus family).

\begin{table}[htbp]
\centering
\scriptsize
\renewcommand{\arraystretch}{0.9}
\caption{In-domain reference: calibration and test drawn from the same corpus family (1\% budget, mean of five seeds).}
\label{tab:in_domain}
\begin{tabular}{lcccc}
\toprule
 & \multicolumn{4}{c}{FPR (\%) / recall (\%) at a 1\% budget} \\
\cmidrule(lr){2-5}
Corpus (train = test family) & TF-IDF LR & Emb.\ LR & Two-class $k$NN & CyberCane \\
\midrule
Nazario + SpamAssassin & 0.9 / 99.9 & 1.2 / 99.9 & 1.1 / 99.9 & 1.1 / 98.0 \\
DataPhish 2025 & 0.8 / 85.7 & 1.6 / 93.6 & 1.0 / 91.6 & 1.4 / 87.0 \\
\bottomrule
\end{tabular}

\end{table}

\begin{table}[htbp]
\centering
\scriptsize
\renewcommand{\arraystretch}{0.9}
\caption{Preregistered experiments and outcomes.}
\label{tab:prereg}
\begin{tabular}{llll}
\toprule
Experiment & Question & Criterion & Outcome \\
\midrule
Pilot and replication & FPR overrun under benign shift & CyberCane below B3 and B9 at $\geq$2/3 budgets, CIs separated & Met; replicated over 5 seeds \\
Decontamination & Robustness to cross-corpus overlap & (robustness check) & FPR effect unchanged; recall corrected \\
Calibration sources & Generality across benign sources & Criterion met in $\geq$2/3 sources & Met, 3/3 \\
Attack-style shift & Human $\rightarrow$ LLM-written phishing & $\geq$4/5 seeds, CIs separated & Not met (no method overruns) \\
Zero-shot LLM & Comparator & none & Holds budget at lower recall \\
Second zero-shot LLM & Strict-end ranking vs.\ Llama 3.3 70B & CyberCane ahead at 1\% realized FPR, $\geq$4/5 seeds, $\geq$2/3 sources & Not met; the 70B LLM ranks higher \\
Matched-FPR frontier & Ranking vs.\ calibration (oracle) & none (descriptive) & Frontiers cross; trained baselines fail by calibration \\
Label-free recalibration & Can unlabeled mail repair thresholds? & CyberCane below B3 and B9 under the mixture rule, $\geq$4/5 seeds, $\geq$2/3 sources & Met (DataPhish); does not generalize to older-corpus shifts \\
Legitimate-mail-only shift & Generality beyond DataPhish & Overrun in $\geq$2/6 pairs, CyberCane better in each, $\geq$4/5 seeds & Met (4/6 pairs with overrun, 5/5 seeds each) \\
Score fusion & CyberCane $\times$ LLM & Recall $+3$ pts at no higher FPR, $\geq$4/5 seeds & Not met (3/5); not confirmed out of sample; with Llama 3.3 70B also not met (1/3) \\
CSLS correction & Label-free adaptation & FPR halved, recall loss $\leq$5 pts, $\geq$2/3 sources & Not met (1/3) \\
Few-shot adaptation & Recall gain without retraining & Recall $+5$ pts at $k{=}25$, $\geq$2/3 sources & Not met (0/3) \\
LLM verification & Retrieval-gated LLM cascade & FPR $\leq$2\% in 3/3 sources & Not met (\NumVoneFPRLo--\NumVoneFPRHi); with Llama 3.3 70B, recall within 3 pts of the LLM in 1/3 \\
Local embedder & Embedder dependence & Calibration effect; best pAUC$_{\leq 2\%}$; no material loss & Calibration met (3/3); ranking and loss not met \\
Score-weight sensitivity & Dependence on hand-set weights & Calibration effect in all \NumWoneConfigs{} settings & Met (\NumWoneConfigs/\NumWoneConfigs, 5/5 seeds each) \\
Explanation audit (E1) & Explanation groundedness & Support lower bound $\geq$0.80; contradiction upper bound $\leq$0.10; checker precision $\geq$0.90 & Not met; not met (largely annotation error, post hoc); met \\
Natural-identifier audit (TB) & Redaction of natural identifiers & Personal-leak upper bound $\leq$0.10; harmful masking upper bound $\leq$0.05 & Not met (mostly label artifacts); met \\
Planted-identifier audit & Redaction recall & Structured types $\geq$99\% in both contexts & Not met by v2.2 (day-first dates, IP); met post hoc by v2.3 \\
Padding robustness & Benign padding, whole-email scoring & Retention $\geq$ trained detectors, $\geq$2/3 sources & Not met (0/3) \\
Segment-level scoring & Defense against padding & Retention $\geq$ trained detectors; clean recall within 2 pts; $\geq$2/3 sources & Robustness met (2/3); clean recall not met (1/3) \\
Authentication checks & Live-DNS false alarms & $\geq$30\% fewer; $\leq$2 phishing lost & Met (\NumAuthStrict{} $\rightarrow$ \NumAuthRfc{}; none lost) \\
Cost and latency & Deployment cost & none (descriptive) & Table~\ref{tab:llm_compare} \\
\bottomrule
\end{tabular}
\end{table}

\section{Human Annotation Protocol}\label{app:annotation}

Two preregistered audits (Appendix~\ref{app:prereg}) use a human annotator: the explanation audit (E1) and an audit of redaction on naturally occurring identifiers (TB), which complements the planted-identifier audit (Table~\ref{tab:privacy_planted}). One annotator, who is one of the authors, labels both, so inter-annotator agreement is not reported.

\textbf{Sample.} E1 draws test emails by outcome cell (true positive, false positive, true negative, review), and each email is explained twice, with and without ontology context. TB draws test emails stratified by corpus and label. All items show text after redaction. To limit workload, the planned sample (120 E1 and 50 TB emails) was released in two stages, the second conditional on a quality gate: at most 15\% errors on sentences whose truth follows mechanically from the evidence panel, over the first 30 stage-1 E1 items. The gate was passed (8 errors on 68 sentences), but stage~2 was not released, so the analyzed sample is stage~1: 60 E1 emails (120 explanations, half of each outcome cell) and 30 TB emails.

\textbf{Task.} In E1, the annotator sees the email, the verdict, and the evidence panel (Phase~1 indicators, sender checks, retrieved neighbors with similarities, ontology classes), labels each explanation sentence as \emph{supported}, \emph{contradicted}, \emph{unverifiable}, or \emph{no claim}, and rates helpfulness and consistency with the verdict. In TB, each redaction placeholder is shown with the text it replaced and classified as a personal identifier, a company identifier, or a harmless word (with whether hiding it harms judgment); identifiers left visible are marked. The annotator is not told that two explanation variants exist, although, as an author, the annotator may know the design; the two explanations of an email are at least 20 positions apart, and neither the checker's labels nor the email's true label is shown. Written guidelines, a short tutorial, and practice items with feedback precede the task; practice answers are not analyzed.

\textbf{Analysis.} Units are sentences (E1) and identifiers (TB); 95\% confidence intervals come from a cluster bootstrap over emails (10{,}000 resamples). Wording thresholds were fixed in advance: explanations are ``largely grounded'' only if the lower bound of the support rate is at least 0.80 and ``rarely wrong'' only if the upper bound of the contradiction rate is at most 0.10; personal identifiers are ``largely removed'' only if the upper bound of their leak rate is at most 0.10.

\textbf{Deviations.} One annotator instead of two; a revised TB form and, for E1, in-tool reminders and one more practice item, both after the pilot showed systematic labeling errors; the staged release; and counting four fully answered TB emails that the annotator had set aside as complete. Each was recorded before any affected result was computed. The mechanical check of sentences that restate the evidence panel (Section~\ref{sec:system}) is post hoc. Guidelines, item files, and anonymized labels will be released with the code.

\clearpage\label{sec:backmatter-start}\section*{Author information}
\noindent University of Maryland, Baltimore County; Bowie State University
\par\medskip
\noindent Safayat Bin Hakim: \href{mailto:shakim3@umbc.edu}{shakim3@umbc.edu}\\
Aniqa Afzal: \href{mailto:aafz1@umbc.edu}{aafz1@umbc.edu}\\
Qi Zhao: \href{mailto:qiz1@umbc.edu}{qiz1@umbc.edu}\\
Vigna Majmundar: \href{mailto:vignam1@umbc.edu}{vignam1@umbc.edu}\\
Pawel Sloboda: \href{mailto:slobodap1024@students.bowiestate.edu}{slobodap1024@students.bowiestate.edu}\\
Houbing Herbert Song: \href{mailto:songh@umbc.edu}{songh@umbc.edu}
\par\medskip
\setlength{\bibsep}{1pt}
\bibliographystyle{unsrtnat}
\bibliography{references}

\end{document}